\documentclass{article}
\usepackage{bsymb}
\usepackage{url}
\usepackage{stmaryrd}
\usepackage{amsthm}
\usepackage{amsmath,amssymb}
\newcommand{\sqlerc}{\textsf{\mbox{sql2erc}}}
\newcommand{\sqlerca}{\textsf{\mbox{sql2erc$_A$}}}
\newcommand{\sqlerco}{\textsf{\mbox{sql2erc$_O$}}}
\newcommand{\sqlercas}{\textsf{\mbox{sql2erc$_{AS}$}}}

\newcommand{\ebsql}{\textsf{\mbox{EB2SQL}}}
\newcommand{\ebsqla}{\textsf{\mbox{EB2SQL$_A$}}}
\newcommand{\ebsqlo}{\textsf{\mbox{EB2SQL$_O$}}}
\newcommand{\ebsqlas}{\textsf{\mbox{EB2SQL$_{AS}$}}}
\newcommand{\ebsqlr}{\textsf{\mbox{EB2SQL$_{RES}$}}}
\newcommand{\ebsqlos}{\textsf{\mbox{EB2SQL$_{OS}$}}}
\newcommand{\rep}{\mbox{$\mathit rep$}}
\newcommand{\repe}{\mbox{$\mathit rep_E$}}
\newcommand{\eb}{\textsf{\mbox{eb}}}
\newcommand{\eba}{\textsf{\mbox{eb$_A$}}}
\newcommand{\ebas}{\textsf{\mbox{eb$_{AS}$}}}
\newcommand{\lb}{\llbracket}
\newcommand{\rb}{\rrbracket}
\newtheorem{thm}{Theorem}
\newtheorem{case}{Case}
\makeatletter
\@addtoreset{case}{thm}
\makeatother
\newcommand{\sql}[1]{\mbox{\texttt{#1}}}
\begin{document}
\title{Formal Semantics and Soundness of a Translation from Event-B
  Actions to SQL Statements}
\author{
 Tim Wahls\\
Department of Mathematics and Computer Science\\
Dickinson College\\
wahlst@dickinson.edu
}
\maketitle
 
 \begin{abstract}
The EventB2SQL tool translates Event-B models to persistent
Java applications that store the state of the model in a relational database.
Most Event-B assignments are translated directly to SQL database
modification statements, which can then
be executed against the database.  In this work, we present a formal semantics 
for and prove the soundness of the translation of sets of assignment
statements representing the actions of an Event-B event.
This allows the generated code to be used with
confidence in its correctness.
 \end{abstract}


\section{Introduction}
  The EventB2SQL 
 tool\footnote{\url{https://sourceforge.net/projects/eventb2sql/}}~\cite{wahls2014}
 translates Event-B~\cite{Abrial2010} models to 
 Java applications that store the state of the model in a relational 
 database (either MySQL or SQLite).  This makes the application persistent.  
 Each event is translated as a database transaction, so the generated code is
 readily used in multi-threaded and client-server applications.
 Carrier sets are translated as Java generic type parameters
 and elements of those sets
 are stored in the database, allowing the code generated from the model to 
 manipulate almost any Java object.
 EventB2SQL has been used in the re-development of an Enterprise 
 Resource Planning system~\cite{Wahls2015a} and the development of an
 Android application for checking medication interactions and 
 contraindications~\cite{Wahls2016}.

  Soundness is particularly critical for code generators for Event-B, as
 Event-B models are typically verified to satisfy correctness and safety
 properties.
 If the code generated from such models does not satisfy those same 
 properties, the code generation tool is worse than useless and may even 
 present a danger to persons or property.
 Most Event-B code generators~\cite{Mery:2011,EB10,EB11,Wright09} translate
 concrete models that are already relatively close to code, so that the
 correspondence between model and code is relatively straightforward. This
 may partially explain why no proof of the the soundness of the translation
 performed by most of these tools has been done.
 The refinement proofs performed as the abstract model is refined to a
 concrete one suitable for translation in some sense stand in for a
 soundness proof for the code generation algorithm.
 EventB2SQL is different in that it is meant to translate relatively abstract
 Event-B machines that use features such as simultaneous assignment,
 quantifiers, set comprehensions and variables of set and relation types.
 Hence, the conceptual distance between the translation source and target
 is much greater, and the correctness of that translation much less readily
 apparent.  Refinement chains when using EventB2SQL are short or even
 non-existent.  This is a tremendous advantange in terms of the time and
 effort needed to reach executable code, but the relationship between the
 abstract model and final implementation is much less apparent to the 
 user.  Hence, assuring the user of the soundness of the code generated by
 EventB2SQL is critical.
 
  EventB2SQL translates most Event-B assignment statements directly to SQL
 \sql{insert} and \sql{delete} statements, which are then
 executed against the model state as represented in the database.  
 In this work, we
 give a formal semantics for and prove the soundness of this part of the
 translation, thus allowing the code generated by EventB2SQL to be used
 with confidence in its correctness.  We first give formal semantics for
 SQL and Event-B, as adapted from appropriate sources.
 We then show the soundness of the
 translation of Event-B expressions to SQL queries (Section~\ref{sect-expr}),
 and finally the soundness of the translation of a set of
 Event-B assignment statements (representing the body of an event) to SQL
 \sql{insert} and \sql{delete} statements (Section~\ref{sect-multassign}).

\section{Semantic Definitions}

\begin{figure*}
\centerline{
\begin{tabular}{lcl}
$\sqlerc \lb \sql{r}, db \rb $ &=& $db(\sql{r})$\\
$\sqlerc \lb \sql{select } \tau_1, \ldots , \tau_n$ &=& \{res$_1$, \ldots , res$_n$ $~|~ 
 \mbox{res}_1 : d_1 \# \wedge \ldots \wedge \mbox{res}_n : d_n \# \wedge $\\
~~~~~~~~~ \sql{from} $ r_1  s_1, \ldots ,  r_n  s_n $ &  & 
~~~ $\exists s_1 : \sqlerc \lb r_1, db \rb \cdot \ldots \exists s_m : \sqlerc \lb r_m, db \rb \cdot $\\
~~~~~~~~~ \sql{where} $\varphi, db \rb$
 & & ~~~~~  $\mbox{res}_1 = \sqlerc \lb \tau_1, db \rb \wedge \ldots \wedge$ \\
 & & ~~~~~  $\mbox{res}_n = \sqlerc \lb \tau_n, db \rb$
  $\wedge~ \sqlerc \lb \varphi, db \rb \}$\\
$\sqlerc \lb s.A, db \rb$ &=& $A(s)$\\
$\sqlerc \lb \sql{not } \varphi, db \rb$ &=& $(\neg \sqlerc \lb \varphi, db \rb )$\\
$\sqlerc \lb  \varphi_1 \sql{ and } \varphi_2, db \rb$ &=&
  $ \sqlerc \lb \varphi_1, db \rb \wedge \sqlerc \lb \varphi_2, db \rb$\\
$\sqlerc \lb  \varphi_1 \sql{ or } \varphi_2, db \rb$ &=&
  $ \sqlerc \lb \varphi_1, db \rb \vee \sqlerc \lb \varphi_2, db \rb$\\
$\sqlerc \lb  \tau_1 ~\omega~ \tau_2, db \rb$ &=&
  $ \sqlerc \lb \tau_1, db \rb ~\omega~ \sqlerc \lb \tau_2, db \rb$\\
$\sqlerc \lb \tau \sql{ in } (\sql{select } s_i.A $ &=&
  $\exists s_i : \sqlerc \lb r_i, db \rb \cdot ((\exists s_1 : \sqlerc \lb r_1, db \rb \ldots $\\
~~~~~~~~~~~~ $\sql{from } r_1 s_1, \ldots , r_m s_m$ &  &
 ~~~$ \exists s_{i - 1} : \sqlerc \lb r_{i - 1}, db \rb \exists s_{i + 1} : \sqlerc \lb r_{i + 1}, db \rb
  \ldots $ \\
~~~~~~~~~~~~ $ \sql{where } \varphi), db \rb$ && ~~~ $\exists s_m : \ebsql \lb r_m, db \rb) \cdot 
 \sqlerc \lb \varphi, db \rb )$ \\
 &&  ~~~~~~$\wedge \sqlerc \lb \tau = s_i.A, db \rb $\\
 $\sqlerc \lb \sql{select } \tau_1, \ldots , \tau_n$ &=& \{res$_1$, \ldots , res$_n$ $~|~ 
 \mbox{res}_1 : d_1 \# \wedge \ldots \wedge \mbox{res}_n : d_n \# \wedge $\\
~~~~~~~~~ \sql{from} $r_1  s_1, \ldots,  r_n  s_n$ &  & 
~~~ $\exists s_1 : \sqlerc \lb r_1, db \rb \cdot \ldots \exists s_m : \sqlerc \lb r_m, db \rb \cdot $\\
~~~~~~~~~ \sql{where} $\varphi$
 & & ~~~~~  $\mbox{res}_1 = \sqlerc \lb \tau_1, db \rb \wedge \ldots
  \wedge \mbox{res}_n = \sqlerc \lb \tau_n, db \rb \wedge $ \\
~$\sql{union select } \tau_1', \ldots , \tau_n'$ & &
   ~~~~~  $ \sqlerc \lb \varphi, db \rb $\\
~~~~~~~~~ \sql{from} $r_1' s_1', \ldots, r_n'  s_n'$  & &  
~ $\vee \exists s_1' : \sqlerc \lb r_1', db \rb \cdot \ldots \exists s_m' : 
 \sqlerc \lb r_m', db \rb \cdot $ \\
~~~~~~~~~ \sql{where} $\varphi', db \rb $ &  & 
 ~~~~~  $\mbox{res}_1 = \sqlerc \lb \tau_1', db \rb \wedge \ldots
  \wedge \mbox{res}_n = \sqlerc \lb \tau_n', db \rb \wedge $ \\
 & & ~~~~~ $ \sqlerc \lb \varphi', db \rb \}$\\
$\sqlerc \lb \sql{select count}(s.A)$ $\sql{from } r  s, db \rb $ &=&
$\mathit{CNT} \{A(s) ~|~ s : \sqlerc \lb r, db \rb \} $\\
\end{tabular}
}
\caption{The definition of the \sqlerc\ operator.}
\label{fig-sql2erc}
\end{figure*}
  
\subsection{Formal Semantics of SQL}

  We construct a formal semantics for SQL by combining the translation of SQL queries
 to Entity-Relationship Calculus (ERC) given in~\cite{sql1994} with the denotational-style
 semantics of \sql{insert} and \sql{delete} statements given in~\cite{Meira1990}.
 Following~\cite{Meira1990}, we define a database $db$ as a function from relation names to
 relations.
 A relation is a set of tuples with named attributes.  We then modify the \sqlerc\ operator
 of~\cite{sql1994} from a syntax transformer to a semantic function that takes an SQL
 query and a database as its arguments, and returns a relation as its result
 (Figure~\ref{fig-sql2erc}).  When the returned relation has a single tuple and a single
 attribute, we often treat it as an atomic (primitive) value.
 In the semantics, r is a relation name, $\tau_i$ represents a term,
 $d_i \#$ the set of all values in the database of the type of term $\tau_i$, and
 $\varphi_i$ a boolean valued formula.  $s_i$ is a tuple variable ranging over
 relation valued expressions
 $r_i$, and tuple variables are unique within a query.  $A$ is an attribute, and $A(s)$
 is the value of attribute $A$ for tuple variable $s$.  Finally, $:$ is used for set
 membership, and  $\omega$ is a binary operator.  The semantics assumes that no
 tables contain duplicate elements, and that no queries produce duplicates.
 All queries produced by EventB2SQL that could potentially have duplicates
 in their results
 use the keyword distinct to eliminate such duplicates.  Where needed,
 justification
 that queries do not produce duplicates is provided in the following proofs.
  
 Beyond the adaptation to a denotational style, the definition of \sqlerc\
 given in Figure~\ref{fig-sql2erc} differs from the definition
given in~\cite{sql1994} in two significant ways:
\begin{itemize}
\item the \sqlerc\ operator is applied to relation valued expressions
 $r_i$ appearing in the \sql{from} clause
 of an SQL query.  This was not done in~\cite{sql1994}, so the semantics given there
 does not consider derived relations (subqueries nested within the \sql{from} clause of 
 another query).  This may be due to the fact that derived relations were a relatively
 new feature in SQL at the time of that work. 
\item the rule for count is simplified from the rule for the $\sqlerc_{\gamma}$ operator
that translates queries with \sql{group by} and \sql{having} clauses
\end{itemize}

\begin{figure*}
\centerline{
\begin{tabular}{lcl}
$\sqlerca \lb \sql{insert ignore into r } S, db \rb$ &=& 
 $[\sql{r} \rightarrow db(\sql{r}) \bunion \sqlerc \lb S, db \rb]db$ \\
$\sqlerca \lb \sql{delete from r}, db \rb $ &=& $[\sql{r} \rightarrow \{\}]db$\\
$\sqlerca \lb \sql{delete from r}$ &=& $[\sql{r} \rightarrow db(\sql{r})~
  \setminus$ \\
~~~~~~~~~~~~$\sql{where }\varphi, db \rb$ & &
~~~$\sqlerc \lb \sql{select } \mbox{rtmp.}A_1, \ldots , \mbox{rtmp.}A_n$ \\
& & ~~~~~~~~~~~~~\sql{from} r rtmp \\
& & ~~~~~~~~~~~~~\sql{where} [rtmp/r]$\varphi, db \rb]db$ \\
\\
$\sqlerca \lb \lambda, db \rb$ &=& $db$\\ 
$\sqlerca \lb A; AS, db \rb$ &=& $\sqlerca \lb AS, 
  \sqlerca \lb A, db \rb \rb$\\
\end{tabular}
}
\caption{The definition of the \sqlerca\ operator.}
\label{fig-sqlerca}
\end{figure*}

  Next, we introduce a new semantic function \sqlerca\ that defines the semantics of
 \sql{insert} and \sql{delete} statements.  \sqlerca\ takes one or more of
  these statements and
 a database as its arguments, and returns an updated database as its result.  The definitions
 in Figure~\ref{fig-sqlerca} are adapted from the definitions of semantic functions 
 {\it Value\_InSel}, {\it Value\_Del} and {\it Value\_DelCond} in~\cite{Meira1990}.
 In the first rule, $S$ is an SQL query with a return value that is compatible
 with relation r (in terms of number of attributes and attribute types).
 In the second rule for \sql{delete},
 $A_1, \ldots , A_n$ are all of the attributes of \sql{r}.
 The notation $[$r $\rightarrow v]db$ is used for updating functions, and evaluates to a 
 function that is the same as $db$ except that r maps to $v$.  The notation
 [rtmp/r]$\varphi$ evaluates to $\varphi$ with all free occurrences of r replaced by rtmp.
 This is done for compatibility with the notation used in~\cite{sql1994}.
 The final two rules gives the semantics of a sequence of \sql{insert} and 
 \sql{delete} statements, where $\lambda$ is an empty sequence,
 $A$ represents a single statement, and $AS$ a sequence of statements.
 The final rule evaluates $AS$ in the state resulting from evaluating $A$.

\subsection{Formal Semantics of Event-B}
    
\begin{figure*}
\centerline{
\begin{tabular}{lcl}
$\eb \lb \mbox{v}, m \rb$ &=& $m(\mbox{v})$ \\
$\eb \lb \neg b, m \rb$ &=& $\neg \eb \lb b, m \rb$\\
$\eb \lb b_1 \wedge b_2 , m \rb$ &=& $\eb \lb b_1, m \rb \wedge
  \eb \lb b_1, m \rb$\\
$\eb \lb b_1 \vee b_2 , m \rb$ &=& $\eb \lb b_1, m \rb \vee
  \eb \lb b_1, m \rb$\\
$\eb \lb x \in s, m \rb$ &=& $\eb \lb x, m \rb \in \eb \lb s, m \rb$\\
$\eb \lb x \subset s, m \rb$ &=& $\eb \lb x, m \rb \subset \eb \lb s, m \rb$\\
$\eb \lb x \subseteq s, m \rb$ &=& $\eb \lb x, m \rb 
  \subseteq \eb \lb s, m \rb$\\
$\eb \lb s_1 \bunion s_2 , m \rb$ &=& $\{ x ~|~ x \in \eb \lb s_1, m \rb
  \vee x \in \eb \lb s_2, m \rb \}$ \\
$\eb \lb s_1 \binter s_2 , m \rb$ &=& $\{ x ~|~ x \in \eb \lb s_1, m \rb
  \wedge x \in \eb \lb s_2, m \rb \}$ \\
$\eb \lb s_1 \setminus s_2 , m \rb$ &=& $\{ x ~|~ x \in \eb \lb s_1, m \rb
  \wedge x \not \in \eb \lb s_2, m \rb \}$ \\
$\eb \lb \card(s), m \rb$ &=& $|\eb \lb s, m \rb|$\\
$\eb \lb s_1 \cprod s_2, m \rb$ &=& 
$\{ x \mapsto y ~|~ x \in \eb \lb s_1, m \rb \wedge 
 y \in \eb \lb s_2, m \rb \}$\\
$\eb \lb \dom(r), m \rb$ &=& $\{ x ~|~ \exists y \cdot x \mapsto y \in 
  \eb \lb r, m \rb$\}\\
$\eb \lb \ran(r), m \rb$ &=& $\{ y ~|~ \exists x \cdot x \mapsto y \in 
  \eb \lb r, m \rb$\}\\
$\eb \lb s \domres r, m \rb$ &=& $\{ x \mapsto y ~|~ x \mapsto y \in 
  \eb \lb r, m \rb \wedge x \in \eb \lb s, m \rb$\}\\
$\eb \lb s \domsub r, m \rb$ &=& $\{ x \mapsto y ~|~ x \mapsto y \in 
  \eb \lb r, m \rb \wedge x \not \in \eb \lb s, m \rb$\}\\
$\eb \lb r \ranres s, m \rb$ &=& $\{ x \mapsto y ~|~ x \mapsto y \in 
  \eb \lb r, m \rb \wedge y \in \eb \lb s, m \rb$\}\\
$\eb \lb r \ransub s, m \rb$ &=& $\{ x \mapsto y ~|~ x \mapsto y \in 
  \eb \lb r, m \rb \wedge y \not \in \eb \lb s, m \rb$\}\\
$\eb \lb r_1 ; r_2, m \rb$ &=& $\{ x \mapsto y ~|~ \exists z \cdot 
 x \mapsto z \in 
  \eb \lb r_1, m \rb \wedge z \mapsto y \in \eb \lb r_2, m \rb$\}\\
$\eb \lb r_1 \circ r_2, m \rb$ &=& $\eb \lb r_2 ; r_1, m \rb$\\ 
$\eb \lb r_1 \ovl r_2, m \rb$ &=& $\eb \lb r_2 \bunion (\dom(r_2) \domsub r_1),
 m \rb$\\
$\eb \lb r^{-1}, m \rb$ &=& $\{ y \mapsto x ~|~ x \mapsto y \in 
  \eb \lb r, m \rb $\}\\
$\eb \lb r[s], m \rb$ &=& $\{ y ~|~ \exists x \cdot x \in \eb \lb s, m \rb
 \wedge x \mapsto y \in \eb \lb r, m \rb $\}\\
\\
$\eba \lb \mbox{v} \bcmeq E, m \rb $ &=& $[\mbox{v} \rightarrow 
  \eb \lb E, m \rb] m $\\
\\
$\ebas \lb \lambda, m \rb $ &=& $m$\\
$\ebas \lb \mbox{v} \bcmeq E || AS, m \rb $ &=& $[\mbox{v} \rightarrow 
 \eb \lb E, m \rb] \ebas \lb AS, m \rb$\\
\end{tabular}
}
\caption{The definition of the \eb, \eba\ and \ebas\ operators.}
\label{fig-eb}
\end{figure*}

  In the usual denotational style, we
 define the state of an Event-B model
 as a function $m$ mapping Event-B machine variable names
 (identifiers) to values.  The possible value types are integer, boolean,
 sets of integers and booleans, and sets of ordered
 pairs of integers and/or booleans.  Figure~\ref{fig-eb} presents the
 operators \eb\ (defining the semantics of an Event-B expression in a given
 state) and \eba\ (defining the state produced by an assignment).
 The definition of the \eb\ operator is adapted from the semantics of 
 Event-B expressions given in Sect. 3.3
   (Mathematical Notation) of the Rodin User's
 Handbook~\cite{RodinHandbook}.
 In the semantics, v is a machine variable name, $b$ is a boolean expression,
 $x, y$ are set elements,
 $s, s_1, s_2$ are set valued expressions, and $r, r_1, r_2$ are relation
 valued expressions.
 In Event-B, $s \domres r$ returns the pairs in $r$ with key occurring in
 $s$, $s \domsub r$ the pairs in $r$ with key not occurring in $s$, 
 $r \ranres s$ the pairs in $r$ with value occurring in $s$, 
 and $r \ransub s$ the pairs in $r$ with value
 not occurring in $s$.  The $;$ operator is relational composition, 
 $\circ$ is reverse
 composition, $\ovl$ is relational overriding, $^{-1}$ relational inverse,
  and $[]$ relational image of a set of domain elements.
 In the rule for \eba, $E$ is an Event-B expression.

 In the rules for \ebas, $\lambda$ represents 0 assignment statements, $E$ an
 Event-B expression, and $AS$ a set of 0 or more assignments.
 Because Event-B uses simultaneous assignment, the right hand side of each
 assignment is executed in the initial state $m$.  An event is allowed to 
 assign each machine variable at most once, so the order in which the 
 assignments are processed and new values are associated with machine variables
 does not matter.

\subsection{Formal Semantics of the Translation}
    
\begin{figure*}
\centerline{
\begin{tabular}{lcl}
\ebsql(s) & $=$ & \sql{select} stmp.refkey \sql{from} s stmp\\
\ebsql(r) & $=$ & \sql{select} rtmp.id, rtmp.value \sql{from} r rtmp\\
$\ebsql(b_1 \wedge b_2)$ & $=$ & $\ebsql(b_1) \sql{ and } \ebsql(b_2)$\\
$\ebsql(b_1 \vee b_2)$ & $=$ & $\ebsql(b_1) \sql{ or } \ebsql(b_2)$\\
$\ebsql(\neg b)$ & $=$  & $\sql{not } \ebsql(b)$\\
$\ebsql(x = y)$ & $=$ & $\ebsql(x) = \ebsql(y)$\\
$\ebsql(\card(s))$ & = & \sql{select count}(stmp.refkey) \sql{from} \ebsql($s$) stmp\\
$\ebsql(s_1 \bunion s_2)$ &= & \sql{select} s1tmp.refkey \sql{from} \ebsql($s_1$) s1tmp
 \sql{union} \\
& &  \sql{select} s2tmp.refkey \sql{from} \ebsql($s_2$) s2tmp\\
$\ebsql(s_1 \binter s_2)$ & = &\sql{select} s1tmp.refkey \sql{from} \ebsql($s_1$) s1tmp, \\
&&  \ebsql($s_2$) s2tmp \sql{where} s1tmp.refkey = s2tmp.refkey\\
$\ebsql(s_1 \cprod s_2)$ & = & \sql{select} s1tmp.refkey, s2tmp.refkey
  \sql{from} \ebsql($s_1$) s1tmp, \ebsql($s_2$) s2tmp\\
$\ebsql(s_1 \setminus s_2)$ & = & \sql{select} s1tmp.refkey 
 \sql{from} \ebsql($s_1$) s1tmp\\
&& \sql{where} s1tmp.refkey \sql{not in}
(\sql{select} s2tmp.refkey \sql{from} \ebsql($s_2$) s2tmp)\\
$\ebsql(s_1 = s_2)$ & $=$ & $\ebsql(|s_1 \binter s_2| = |s_1| \wedge |s_1| = |s_2|)$\\
$\ebsql(s_1 \subseteq s_2)$ & $=$ & $\ebsql(|s_1 \binter s_2| = |s_1|)$\\
$\ebsql(s_1 \subset s_2)$ & $=$ & $
  \ebsql(|s_1 \binter s_2| = |s_1| \wedge |s_1| \neq |s_2|)$\\
$\ebsql(x \in s) $ & $=$ & $\ebsql(\{x\} \subseteq s)$\\
$\ebsql(\dom(r))$ &=& \sql{select distinct} rtmp.id \sql{from} \ebsql($r$) rtmp\\
$\ebsql(\ran(r))$ &=& \sql{select distinct} rtmp.value \sql{from} \ebsql($r$) rtmp\\
$\ebsql(s \domres r)$ &=& \sql{select} rtmp.id, rtmp.value \sql{from} \ebsql(r) rtmp,
  \ebsql(s) stmp\\
& &  \sql{where} rtmp.id = stmp.refkey\\
$\ebsql(s \domsub r)$ & = & \sql{select} rtmp.id, rtmp.value \sql{from} \ebsql($r$) rtmp\\
&& \sql{where} rtmp.id \sql{not in}
 (\sql{select} stmp.refkey \sql{from} \ebsql($s$) stmp)\\
$\ebsql(r \ranres s)$ &=& \sql{select} rtmp.id, rtmp.value \sql{from} \ebsql(r) rtmp, \\
&& \ebsql(s) stmp \sql{where} rtmp.value = stmp.refkey\\
$\ebsql(r \ransub s)$ & = & \sql{select} rtmp.id, rtmp.value \sql{from} \ebsql($r$) rtmp\\
&& \sql{where} rtmp.value \sql{not in}
 (\sql{select} stmp.refkey \sql{from} \ebsql($s$) stmp)\\
$\ebsql(r_1;r_2)$ &=& \sql{select distinct} r1tmp.id, r2tmp.value
  \sql{from} \ebsql($r_1$) r1tmp, \ebsql($r_2$) r2tmp \\
&& \sql{where} r1tmp.value = r2tmp.id\\
$\ebsql(r_1 \circ r_2)$ &=& $\ebsql(r_2;r_1)$\\
$\ebsql(r_1 \ovl r_2)$ &=& $\ebsql(r_2 \bunion (\dom(r_2) \domsub r_1))$\\
$\ebsql(r^{-1})$ &=& \sql{select} rtmp.value, rtmp.id \sql{from} \ebsql($r$) rtmp\\
$\ebsql(r[s])$ &=& \sql{select distinct} rtmp.value \sql{from} \ebsql(r) rtmp,
 \ebsql(s) stmp \\
& & \sql{where} rtmp.id = stmp.refkey\\
\end{tabular}
}
\caption{The definition of the \ebsql\ operator.}
\label{fig-eb2sql}
\end{figure*}
 
   Figure~\ref{fig-eb2sql} gives the definition of the \ebsql\ operator,
 which translates from Event-B expressions and predicates to SQL
 queries.  \ebsql\ is defined inductively.  It is
 a purely syntactic transformation, so
 no machine state is needed at this point.  In the definition, s is an Event-B
 machine variable of a set type, r is a relation variable, 
 $b$, $b_1$ and $b_2$ are 
  Event-B predicates; $x$ and $y$
  are set elements; and $r$, $r_1$ and $r_2$ are relation valued expressions;
  $f$ is a function valued expression; 
  and $s$, $s_1$ and $s_2$ are
  set valued expressions that are not relations.  Sets are represented as database
  tables with a single column called refkey, while relations are represented as tables
  with two columns: id and value.  Hence, an operation such as $\dom(r)$ which is
  defined in Event-B as $\{x ~|~ \exists y \cdot x \mapsto y \in r\}$ becomes:\\
  \sql{select distinct} rtmp.id \sql{from} \ebsql($r$) rtmp\\
  in SQL.  Tuple variables stmp, s1tmp, s2tmp, rtmp, r1tmp and r2tmp represent
  a fresh variable name in each rule application, so no scoping rules are
  needed.

   The use of some basic set theory in the rules for the $=$, $\subseteq$, $\subset$ and
 $\in$ operators precisely captures the implementation of these operators in the
 EventB2SQL tool.  MySQL does not define the ANSI-standard \sql{intersect} and \sql{except}
 operators, so the query for $s_1 \binter s_2$ selects precisely the elements of $s_1$ that
 also occur in $s_2$, and for $s_1 \setminus s_2$ the elements of $s_1$ that do not
 occur in $s_2$.  
    
\begin{figure*}
\centerline{
\begin{tabular}{lcl}
$\ebsqla(\mbox{s} \bcmeq \mbox{s} \bunion s_1)$ & $=$ & 
  \sql{insert ignore into} s \sql{select} stmp.refkey \sql{from} s' stmp\\
$\ebsqla(\mbox{s} \bcmeq \mbox{s} \setminus s_1)$ & $=$ & 
\sql{delete from} s 
 \sql{where} s.refkey \sql{in }\\
&& ~~~\sql{(select} \mbox{s1tmp.refkey} \sql{from} \mbox{s' s1tmp})\\
$\ebsqla(\mbox{s} \bcmeq \mbox{s} \binter s_1$) & $=$ & 
 \sql{delete from} s \sql{where} s.refkey \sql{in} \\
 && ~~~(\sql{select} \mbox{s1tmp.refkey} \sql{from} \mbox{s' s1tmp})\\
$\ebsqla(\mbox{r} \bcmeq \mbox{r} \ovl r_1)$ &=& 
 \sql{delete from} r \sql {where} r.id \sql{in}\\
 && ~~~
 (\sql{select} r1tmp.id \sql{from} r' r1tmp); \\
 && \sql{insert ignore into} r \sql{select} r2tmp.id, r2tmp.value\\
 && ~~~~~~~~~~~~~~~~~~~~~~~~~~~~~~\enskip\sql{from} r' r2tmp\\
$\ebsqla(\mbox{r} \bcmeq s_1 \domsub \mbox{r})$ &=& 
 \sql{delete from} r \sql {where} r.id \sql{in}\\
 && ~~~
 (\sql{select} stmp.refkey \sql{from} r' stmp) \\
$\ebsqla(\mbox{r} \bcmeq s_1 \domres \mbox{r})$ &=& 
 \sql{delete from} r \sql {where} r.id \sql{in}\\
 && ~~~
 (\sql{select} stmp.refkey \sql{from} r' stmp) \\
$\ebsqla(\mbox{r} \bcmeq \mbox{r} \ransub s_1 $) &=& 
 \sql{delete from} r \sql {where} r.value
  \sql{in}\\
 && ~~~
 (\sql{select} stmp.refkey \sql{from} r' stmp) \\
$\ebsqla(\mbox{r} \bcmeq \mbox{r} \ranres s_1 )$ &=& 
 \sql{delete from} r \sql {where} r.value \sql{in}\\
 && ~~~
 (\sql{select} stmp.refkey \sql{from} r' stmp) \\
$\ebsqla(\mbox{s} \bcmeq s_1$) & $=$ & 
\sql{delete from} s;\\
&& \sql{insert ignore into} s
 \sql{select} s1tmp.refkey \\
 && ~~~~~~~~~~~~~~~~~~~~~~~~~~~~~~\enskip\sql{from} s' s1tmp\\
$\ebsqla(\mbox{r} \bcmeq r_1$) & $=$ & 
\sql{delete from} r;\\
&& \sql{insert ignore into} r
 \sql{select} r1tmp.id, r1tmp.value \\
 && ~~~~~~~~~~~~~~~~~~~~~~~~~~~~~~\enskip\sql{from} r' r1tmp\\
\\
$\ebsqlas \lb \lambda, db \rb$ & $=$ & $db$\\
$\ebsqlas \lb v \bcmeq E || AS, db \rb$ & $=$ &  
 $\ebsqlas \lb AS,$ \\
&& ~~~
 $[v' \rightarrow \mathit{undef}]
  \sqlerca \lb \ebsqla(v \bcmeq E), db \rb \rb$\\
\\
$\ebsqlo \lb \mbox{s} \bcmeq \mbox{s} \bunion s_1, db \rb$ & $=$ & 
 $[\mbox{s}' \rightarrow \sqlerc \lb \ebsql(s_1), db \rb]db$\\
$\ebsqlo \lb \mbox{s} \bcmeq \mbox{s} \setminus s_1, db \rb$ & $=$ & 
 $[\mbox{s}' \rightarrow \sqlerc \lb \ebsql(s_1), db \rb]db$\\
$\ebsqlo \lb \mbox{s} \bcmeq \mbox{s} \binter s_1, db \rb$ & $=$ & 
 $[\mbox{s}' \rightarrow \sqlerc \lb \ebsql(\mbox{s} \setminus s_1), db \rb]db$\\
$\ebsqlo \lb \mbox{r} \bcmeq \mbox{r} \ovl r_1, db \rb$ &=& 
 $[\mbox{r}' \rightarrow \sqlerc \lb \ebsql(r_1), db \rb]db$\\
$\ebsqlo \lb \mbox{r} \bcmeq s_1 \domsub \mbox{r}, db \rb$ &=& 
 $[\mbox{r}' \rightarrow \sqlerc \lb \ebsql(s_1), db \rb]db$\\
$\ebsqlo \lb \mbox{r} \bcmeq s_1 \domres \mbox{r}, db \rb$ &=& 
 $[\mbox{r}' \rightarrow \sqlerc \lb \ebsql(\dom(\mbox{r}) 
 \setminus s_1), db \rb]db$\\
$\ebsqlo \lb \mbox{r} \bcmeq \mbox{r} \ransub s_1, db \rb$ &=& 
 $[\mbox{r}' \rightarrow \sqlerc \lb \ebsql(s_1), db \rb]db$\\
$\ebsqlo \lb \mbox{r} \bcmeq \mbox{r} \ranres s_1, db \rb$ &=& 
 $[\mbox{r}' \rightarrow \sqlerc \lb \ebsql(\ran(\mbox{r})
 \setminus s_1), db \rb]db$\\
$\ebsqlo \lb \mbox{s} \bcmeq s_1, db \rb$ & $=$ & 
 $[\mbox{s}' \rightarrow \sqlerc \lb \ebsql(s_1), db \rb]db$\\
$\ebsqlo \lb \mbox{r} \bcmeq r_1, db \rb$ & $=$ & 
 $[\mbox{r}' \rightarrow \sqlerc \lb \ebsql(r_1), db \rb]db$\\
\\
$\ebsqlos \lb \lambda, db \rb$ & $=$ & $db$\\
$\ebsqlos \lb A || AS, db \rb$ & $=$ &  
 $\ebsqlos \lb AS, \ebsqlo \lb A, db \rb \rb$\\
\\
$\ebsqlr \lb AS, db \rb$ & $=$ & 
  $\ebsqlas \lb AS, \ebsqlos \lb AS, db \rb \rb$\\
\end{tabular}
}
\caption{The definition of the \ebsqla, \ebsqlas,  \ebsqlo,
 \ebsqlos\  and \ebsqlr\ operators.}
\label{fig-eb2sqla}
\end{figure*}

The \ebsqla\ operation (Figure~\ref{fig-eb2sqla}) translates Event-B
assignments into SQL \sql{insert} and \sql{delete} statements.  The 
translation is again purely syntactic, so no state parameter is required.
In cases where multiple statements are needed to translate one assignment,
they are separated by semicolons.  
tmp is assumed to be a fresh (unused) table name.
The \ebsqlr\ operator (defined at the end of Figure~\ref{fig-eb2sqla})
uses the \sqlerco\ operator (also defined in Figure~\ref{fig-eb2sqla})
to evaluate an expression derived from the right hand side of the
assignment and bind that result to the primed identifier (either s' or
r') in the state used to evaluate the query.
The first eight rules capture the 
optimizations introduced in~\cite{Wahls2015a}.  The ninth rule could be
used for an assignment where any of the first three rules could be used,
and the tenth rule could be used where any of rules four through eight
could be used.  This is an issue for efficiency only, not correctness, as we
will prove soundness no matter which of the applicable rules is used.
The implementation of the EventB2SQL tool always uses the first applicable
rule.
The \ebsqlas\ operation (also in Figure~\ref{fig-eb2sqla}) translates a
set of Event-B assignment statements into SQL.  $E$ is an expression of a
set or relation type that is compatible with the type of $v$.
As evaluating the result of applying \ebsqla\ defines a primed version of
the name on the left hand side of the assignment in the resulting state,
we use $[v' \rightarrow \mathit{undef}]$ 
to ensure that $v'$ is not defined in the overall result of evaluating
the collection of Event-B assignments.
As shown in the proof of Theorem~\ref{theorem-ass}, this technique for
evaluating a collection of assignments preserves the simultaneous assignment
semantics of Event-B, so the order that the assignments are evaluated in
does not matter.

The \ebsqlo\ operator (still in Figure~\ref{fig-eb2sqla}) binds a primed
version of the name on the left hand side of an assignment to a value 
derived from the expression on the right hand side.  This value is then
used in evaluating the query that \ebsqla\ generates for the assignment.
The effects of using \ebsqla\ and \ebsqlo\ together are as follows:
\begin{itemize}
\item for $\mbox{s} \bcmeq \mbox{s} \bunion s_1$: insert all values in $s_1$ 
 that do not already occur in s into s
\item for $\mbox{s} \bcmeq \mbox{s} \setminus s_1$: delete all values in 
 $s_1$ from s
\item for $\mbox{s} \bcmeq \mbox{s} \binter s_1$: delete all values in 
 $\mbox{s} \setminus s_1$ from s
\item for $\mbox{r} \bcmeq \mbox{r} \ovl r_1$: delete all pairs in 
 r with key occurring as a key in $r_1$, then insert all pairs in
 $r_1$ into r
\item for $\mbox{r} \bcmeq s_1 \domsub \mbox{r}$: delete all pairs in 
 r with key occurring in $s_1$
\item for $\mbox{r} \bcmeq s_1 \domres \mbox{r}$: delete all pairs in 
 r with key occurring in $\dom(\mbox{r}) \setminus s_1$
\item for $\mbox{r} \bcmeq \mbox{r} \ransub s_1 $: delete all pairs in 
 r with value occurring in $s_1$
\item for $\mbox{r} \bcmeq \mbox{r} \ranres s_1 $: delete all pairs in 
 r with value occurring in $\ran(\mbox{r}) \setminus s_1$
\item for $\mbox{s} \bcmeq s_1$: delete all values from s, then insert all 
 values in $s_1$ into s
\item for $\mbox{r} \bcmeq r_1$: delete all pairs from r, then insert all 
 pairs in $r_1$ into r
\end{itemize}
Finally, \ebsqlos\ returns the state resulting from applying \ebsqlo\ to 
each of a collection of assignments, and \ebsqlr\ uses \ebsqlas\ to evaluate
a collection of assignments in the state resulting from applying \ebsqlos\ 
to that same collection.
   
\section{Proof of Soundness}

\begin{figure*}
\centerline{
\begin{tabular}{lcl}
$\repe(i)$ &=& $i$ \\
$\repe(b)$ &=& $b$ \\
$\repe(\{\})$ &=& $\{\}$ \\
$\repe(\{(\mbox{refkey: } x)\} \bunion S)$ &=& $\{x\} \bunion \repe(S)$\\
$\repe(\{(\mbox{id: } x, \mbox{value: } y)\} \bunion R)$ &=& $\{x \mapsto y\}
 \bunion \repe(R)$\\
\\
$\rep(db)$ &=& $\{\mbox{i} \mapsto \repe(v) ~|~ 
  \mbox{i} \rightarrow v \in db\}$\\
\end{tabular}
}
\caption{The definition of the \rep\ operator that maps database states
 to Event-B states.}
\label{fig-rep}
\end{figure*}

  To show the soundness of the translation, we 
 first define a mapping function to relate database states to Event-B
 model states.  Function \rep\ (Figure~\ref{fig-rep}) uses a helper function
 \repe\ to
 translate SQL values to Event-B values.  It
 removes attribute names from tuples and converts 1-tuples to integers
 or booleans. In Figure~\ref{fig-rep}, $i$ is an integer constant,
 $b$ a boolean constant, $S$ a table representing a set and $R$ a table
 representing a relation or function.
 The role of \rep\ is similar to that of gluing invariants in Event-B
 refinement.

  With  these mechanics in hand, we are finally in a position to state
 the soundness of the translation performed by the EventB2SQL tool as a pair of
 theorems.  The first states the soundness of the translation of expressions
  performed by \ebsql, and is presented in Section~\ref{sect-expr}.
 The second states the soundness of the translation of collections of
 Event-B assignment
 statements, and is given in Section~\ref{sect-multassign}.

\subsection{Proof of Soundness: Translating Expressions}
\label{sect-expr}

\begin{thm}
 For every Event-B expression $E$ that is translated directly to an SQL query
 by the EventB2SQL tool and every correctly typed database state $db$:
  \[\repe(\sqlerc \lb \ebsql (E), db \rb) = \eb \lb E, \rep(db) \rb \] 
\label{theorem}
 \end{thm}

   Hence, the SQL query is equivalent to the Event-B expression that it was 
  translated from.  

      We prove Theorem~\ref{theorem} by structural induction.  The 
 induction takes the form of a case analysis, with one case for each rule 
 for \ebsql\ in Figure~\ref{fig-eb2sql}.
  The second line of each proof case uses the rule,
  and in all but  trivial cases the second
  to last line gives the definition of the operator in Event-B from Sect. 3.3
   (Mathematical Notation) of the Rodin User's
 Handbook~\cite{RodinHandbook}.
  At the base of the induction, an Event-B set (which may be a relation) is
 represented by a database table containing the elements of the set.  As a
 reminder, relations are represented by tables with columns id and value,
 and other sets by tables with
 a single column called refkey.  Predicates of the form res$_i : d_i \#$
 are dropped in the proofs when it is established that res$_i$ is a value in the
 database.  We use subs. to justify a substitution of equals for equals, 
 I.H. for an application of the inductive hypothesis, ERC a step by
 the definition of the Entity-Relationship calculus,
 and the name of an operator to justify a step by definition of that operator.
  
  The definitions and proofs for set operations $\card$, $\bunion$, $\binter$ 
 and $\setminus$  are easily extended
 for relations.  In the relation case, the proofs would have exactly the same 
 structure except that there would be two attributes in the select clause 
 (id and value) rather than the single attribute (refkey) in the non-relation
 case.  Additionally,
 comparisons of the form s1tmp.refkey = s2tmp.refkey would become r1tmp.id =
 r2tmp.id \sql{and} r1tmp.value = r2tmp.value.
 We assume type compatibility of all operands in Event-B expressions, as 
 this is checked by Rodin.
 Several of the less interesting cases are omitted in the interest of space.
 
\begin{case}
$ \repe(\sqlerc \lb \ebsql (\textup{s}), db \rb) = 
 \eb \lb \textup{s}, \rep(db) \rb$
   (basis)
\label{case-set}
\end{case}

 \begin{proof}
 \begin{align*}
 &  \repe(\sqlerc \lb \ebsql ({\mbox s}), db \rb) \\
  &  = \repe(\sqlerc \lb \mbox{\sql{select }stmp.refkey} 
   \sql{ from } \mbox{s stmp}, db \rb) 
  &\text{\ebsql}\\
 & = \repe(\{\mbox{res}_1 ~|~ \mbox{res}_1 :  d_1 \# ~ \wedge 
\exists \mbox{stmp} : \sqlerc \lb \mbox{s}, db \rb \cdot & \\
 & ~~~~~ \mbox{res}_1 = \sqlerc \lb \mbox{stmp.refkey}, db \rb \})
  &\text{\sqlerc}\\
 & = \repe(\{\mbox{res}_1 ~|~ \mbox{res}_1 :  d_1 \# ~ \wedge 
  \exists \mbox{stmp} :  db(\mbox{s}) \cdot & \\
 & ~~~~~~~~ \mbox{res}_1 = \mbox{refkey(stmp)} \rb \} )
  &\text{\sqlerc}\\
 & = \repe(\{\mbox{refkey(stmp)} ~|~ \mbox{stmp} :  db(\mbox{s})\}) &
  \text{subs.} \\
 & = \{x ~|~ x \in  \rep(db)(\mbox{s})\} & \text{\repe} \\
 & = \rep(db)(\mbox{s}) & \rep \\
 & = \eb \lb \mbox{s}, \rep(db) \rb & \eb\ \qedhere
\end{align*}
\end{proof}
 
\begin{case}
$ \repe(\sqlerc \lb \ebsql (\textup{r}), db \rb) \equiv \eb \lb \textup{r},
 \rep(db) \rb $ (basis)
\end{case}
 \begin{proof}
 \begin{align*}
  &  \repe(\sqlerc \lb \ebsql (\mbox{r}), db \rb)  \\
  &  = \repe(\sqlerc \lb \mbox{\sql{select }rtmp.id, rtmp.value } 
   \sql{from } \mbox{r rtmp}, db \rb) 
  &\text{\ebsql}\\
 & = \repe(\{\mbox{res}_1, \mbox{res}_2 ~|~ \mbox{res}_1 :  d_1 \# ~ \wedge 
   \mbox{res}_2 :  d_2 \wedge
\exists \mbox{rtmp} : \sqlerc \lb \mbox{r}, db \rb \cdot & \\
 & ~~~~~ \mbox{res}_1 = \sqlerc \lb \mbox{rtmp.id}, db \rb  
 \wedge \mbox{res}_2 = \sqlerc \lb \mbox{rtmp.value}, db \rb \})
  &\text{\sqlerc}\\
 & = \repe(\{\mbox{res}_1, \mbox{res}_2 ~|~ \mbox{res}_1 :  d_1 \# ~ \wedge 
   \mbox{res}_2 :  d_2 \wedge
\exists \mbox{rtmp} : db(\mbox{r}) \cdot & \\
 & ~~~~~ \mbox{res}_1 = \mbox{id(rtmp)}  
 \wedge \mbox{res}_2 = \mbox{value(rtmp)} \})
  &\text{\sqlerc}\\
 & = \repe(\{\mbox{id(rtmp)}, \mbox{value(rtmp)} ~|~ \mbox{rtmp} : db(\mbox{r}) \})
  &\text{subs.}\\
 & = \{x \mapsto y ~|~ x \mapsto y \in  \rep(db)(\mbox{r}) \}
  &\text{\repe}\\
 & = \rep(db)(\mbox{r}) & \rep\\
 & = \eb \lb \mbox{r}, rep(db) \rb & \text{\eb}\ \qedhere \\
\end{align*}
 \end{proof}
 
\begin{case}
$ \repe(\sqlerc \lb \ebsql (b_1 \wedge b_2), db \rb) \equiv 
 \eb \lb b_1 \wedge b_2, \rep(db) \rb $
\label{case-and}
\end{case}

 \begin{proof}
 \begin{align*}
 &  \repe(\sqlerc \lb \ebsql (b_1 \wedge b_2), db \rb) \\
 &  \equiv \repe(\sqlerc \lb \ebsql(b_1) \sql{ and } \ebsql(b_2), db \rb)
  &\text{\ebsql}\\
&  \equiv \repe(\sqlerc \lb \ebsql(b_1), db \rb 
\wedge \sqlerc \lb \ebsql(b_2), db \rb)
  & \text{\sqlerc}\\
&  \equiv \repe(\sqlerc \lb \ebsql(b_1), db \rb) 
\wedge \repe(\sqlerc \lb \ebsql(b_2), db \rb)
  & \text{\repe}\\
&  \equiv \eb \lb b_1, \rep(db) \rb 
\wedge \eb \lb b_2, \rep(db) \rb 
  & \text{I.H.}\\
&  \equiv \eb \lb b_1 \wedge b_2, \rep(db) \rb &\text{\eb}\ \qedhere \\  
\end{align*}
\end{proof}

\begin{case}
$ \sqlerc \lb \ebsql (b_1 \vee b_2), db \rb \equiv \eb \lb b_1 \vee b_2,
 \rep(db) \rb $
\end{case}

 \begin{proof}
 Symmetric with Case~\ref{case-and}.
\end{proof}

\begin{case}
$ \repe(\sqlerc \lb \ebsql (\neg b), db \rb) \equiv 
 \eb \lb \neg b, \rep(db) \rb $
\label{case-not}
\end{case}

 \begin{proof}
 \begin{align*}
 &  \repe(\sqlerc \lb \ebsql (\neg b), db \rb) \\
 &  \equiv \repe(\sqlerc \lb  \sql{not } \ebsql(b), db \rb)
  &\text{\ebsql}\\
&  \equiv \repe(\neg \sqlerc \lb \ebsql(b), db \rb)
  &\text{\sqlerc}\\
&  \equiv \neg \repe(\sqlerc \lb \ebsql(b), db \rb)
  &\text{\repe}\\
& \equiv \neg \eb \lb b, \rep(db) \rb &\text{I.H.}  \\  
& \equiv \eb \lb \neg b, \rep(db) \rb &\text{\eb}\ \qedhere  \\  
\end{align*}
\end{proof}

\begin{case}
$ \repe(\sqlerc \lb \ebsql (x = y), db \rb) \equiv \eb \lb x = y, \rep(db) \rb $
\label{case-eq}
\end{case}

 \begin{proof}
Symmetric with Case~\ref{case-and}.
\end{proof}

\begin{case}
$ \repe(\sqlerc \lb \ebsql (\card(s)), db \rb) = 
 \eb \lb \card(s), \rep(db) \rb $
\label{case-card}
\end{case}
\begin{proof}
 \begin{align*}
 &  \repe(\sqlerc \lb \ebsql (\card(s)), db \rb) \\
  &  = \repe(\sqlerc \lb \mbox{\sql{select count}(stmp.refkey)} \\
 & ~~~~~ \sql{from } \ebsql (s) \mbox{ stmp}, db \rb )
   &\text{\ebsql}\\
 & = \repe(\mathit{CNT} \{ \mbox{refkey(stmp}) ~|~ 
\mbox{stmp} : 
  \sqlerc \lb \ebsql(s), db \rb \})
 &\text{\sqlerc}\\
 & = \mathit{CNT} \{ x ~|~ x \in \repe(\sqlerc \lb \ebsql(s), db \rb) \}
 &\text{\repe}\\
 & = \mathit{CNT} \{ x ~|~ x \in \eb \lb s, \rep(db) \rb \}
 &\text{I.H.}\\
 & = \mathit{CNT} \eb \lb s, \rep(db) \rb & \text{subs.}\\
 & = | \eb \lb s, \rep(db) \rb| & \text{ERC}\\
 & = \eb \lb \card(s), \rep(db) \rb & \eb\ \qedhere\\
 \end{align*}
\end{proof}

Note that \sql{count} and \sql{count distinct} give the same result when applied
to an operand that does not contain duplicates.  

\begin{case}
$ \repe(\sqlerc \lb \ebsql (s_1 \bunion s_2), db \rb) = 
 \eb \lb s_1 \bunion s_2, \rep(db) \rb $
\label{case-union}
\end{case}

\begin{proof}
\begin{align*}
 &  \repe(\sqlerc \lb \ebsql (s_1 \bunion s_2), db \rb) \\
  &  = \repe(\sqlerc \lb \mbox{\sql{select} s1tmp.refkey} \\
 & ~~~~~ \sql{from } \ebsql (s_1) \mbox{ s1tmp } \sql{union} \\
 & ~~~~~ \mbox{\sql{select} s2tmp.refkey}\\
 & ~~~~~ \sql{from } \ebsql (s_2) \mbox{ s2tmp}, db  \rb )
   &\text{\ebsql}\\
 & = \repe(\{\mbox{res}_1 ~|~ \mbox{res}_1 :  d_1 \#  \wedge (\\
 & ~~~~~ \exists \mbox{s1tmp} : \sqlerc \lb \ebsql (s_1), db \rb \cdot \\
 & ~~~~~~~~ \mbox{res}_1 = \sqlerc \lb \mbox{s1tmp.refkey}, db \rb  \\
 & ~~ \vee \exists \mbox{s2tmp} : \sqlerc \lb \ebsql (s_2), db \rb \cdot \\
 & ~~~~~~~~ \mbox{res}_1 = \sqlerc \lb \mbox{s2tmp.refkey}, db \rb )  \})
  &\text{\sqlerc}\\
 & = \{x ~|~ x \in \repe(\lb \ebsql (s_1), db \rb) \vee
 x \in \repe(\sqlerc \lb \ebsql (s_2), db \rb)\}
  &\text{subs, \repe}\\
 & = \{x ~|~ x \in \eb\lb s_1, \rep(db) \rb \vee
 x \in \eb \lb s_2, \rep(db) \rb\} &\text{I.H.}\\
 & = \eb \lb s_1 \bunion s_2, \rep(db) \rb  & \eb\ \qedhere
 \end{align*}
\end{proof}

Note that the \sql{union} operator in SQL removes duplicates by default.

\begin{case}
$ \repe(\sqlerc \lb \ebsql (s_1 \binter s_2), db \rb 
 = \eb \lb s_1 \binter s_2, \rep(db) \rb $
\label{case-inter}
\end{case}

\begin{proof}
 \begin{align*}
 &  \repe(\sqlerc \lb \ebsql (s_1 \binter s_2), db \rb) \\
  &  = \repe(\sqlerc \lb \mbox{\sql{select} s1tmp.refkey} \\
 & ~~~~~ \sql{from } \ebsql (s_1) \mbox{ s1tmp}, 
 \ebsql(s_2) \mbox{ s2tmp} &
   \\
 & ~~~~~ \mbox{\sql{where} s1tmp.refkey = s2tmp.refkey}, db  \rb) 
   &\text{\ebsql}\\
 & = \repe(\{\mbox{res}_1 ~|~ \mbox{res}_1 :  d_1 \# \wedge \\
 & ~~~~~ \exists \mbox{s1tmp} : \sqlerc \lb \ebsql (s_1), db \rb \cdot \\
 & ~~~~~~~~ \exists \mbox{s2tmp} : \sqlerc \lb \ebsql (s_2), db \rb \cdot \\
 & ~~~~~~~~~~~ \mbox{res}_1 = \sqlerc \lb \mbox{s1tmp.refkey}, db \rb \wedge 
 & \\
 & ~~~~~~~~~~~ \sqlerc \lb \mbox{s1tmp.refkey = s2tmp.refkey}, db  \rb  \})
  &\text{\sqlerc}\\
 & = \repe(\{\mbox{res}_1 ~|~ \mbox{res}_1 :  d_1 \# \wedge \\
 & ~~~~~ \exists \mbox{s1tmp} : \sqlerc \lb \ebsql (s_1), db \rb \cdot \\
 & ~~~~~~~~ \exists \mbox{s2tmp} : \sqlerc \lb \ebsql (s_2), db \rb \cdot \\
 & ~~~~~~~~~~~ \mbox{res}_1 = \mbox{refkey(s1tmp)} \wedge 
  \mbox{refkey(s1tmp) = refkey(s2tmp)} \})
  &\text{\sqlerc}\\
 & = \{x ~|~ x \in \repe( \sqlerc \lb \ebsql (s_1), db \rb) \wedge \\
 & ~~~~~~~~~~ x \in \repe(\sqlerc \lb \ebsql (s_2), db \rb) \}
  &\text{subs., \repe}\\
 \end{align*}
 \begin{align*}
 & = (\{x ~|~ x \in \eb \lb s_1, \rep(db) \rb \wedge 
  x \in \eb \lb s_2, \rep(db) \rb) \})
  &\text{I.H.}\\
 & =  \eb \lb s_1 \binter s2, \rep(db) \rb 
  &\text{\eb\ \qedhere}
 \end{align*}
\end{proof}

Note that if $s_1$ and $s_2$ do not contain duplicates, then each element of
$s_1$ matches at most one element of $s_2$ in the query above.  This ensures
that the query result does not contain duplicates.  MySQL does not have the
ANSI SQL standard \sql{intersect} operator, so \ebsql\ translates $\binter$ as
 shown above.

\begin{case}
$ \repe(\sqlerc \lb \ebsql (s_1 \cprod s_2), db \rb) = \eb \lb s_1 \cprod s_2, \rep(db) \rb $
\end{case}

\begin{proof}
 \begin{align*}
 &  \repe(\sqlerc \lb \ebsql (s_1 \cprod s_2), db \rb) \\
  &  = \repe(\sqlerc \lb \mbox{\sql{select} s1tmp.refkey, s2tmp.refkey} \\
 & ~~~~~ \sql{from } \ebsql (s_1) \mbox{ s1tmp, } \\
 & ~~~~~~~~~~~~\,\ebsql(s_2) \mbox{ s2tmp}, db  \rb) 
   &\text{\ebsql}\\
& = \repe(\{\mbox{res}_1, \mbox{res}_2 ~|~ \mbox{res}_1 :  d_1 \# \wedge 
    \mbox{res}_2 :  d_2 \# \wedge\\
 & ~~~~~ \exists \mbox{s1tmp} : \sqlerc \lb \ebsql (s_1), db \rb \cdot \\ 
 & ~~~~~~~~\exists \mbox{s2tmp} : \sqlerc \lb \ebsql (s_2), db \rb \cdot \\
 & ~~~~~~~~~~~ \mbox{res}_1 = \sqlerc \lb \mbox{s1tmp.refkey}, db \rb \wedge &  \\
 & ~~~~~~~~~~~  \mbox{res}_2 = \sqlerc \lb \mbox{s2tmp.refkey}, db \rb) \}&\text{\sqlerc}\\
 & = \repe(\{\mbox{res}_1, \mbox{res}_2 ~|~ \mbox{res}_1 :  d_1 \# \wedge 
    \mbox{res}_2 :  d_2 \# \wedge\\
 & ~~~~~ \exists \mbox{s1tmp} : \sqlerc \lb \ebsql (s_1), db \rb \cdot \\ 
 & ~~~~~~~~\exists \mbox{s2tmp} : \sqlerc \lb \ebsql (s_2), db \rb \cdot \\
 & ~~~~~~~~~~~ \mbox{res}_1 =  \mbox{refkey(s1tmp)} \wedge 
   \mbox{res}_2 = \mbox{refkey(s2tmp)} \}&\text{\sqlerc}\\
 & = \{x \mapsto y ~|~ x \in \repe(\sqlerc \lb \ebsql (s_1), db \rb) \wedge \\ 
 & ~~~~~~~~~~~~~~~~y \in  \repe(\sqlerc \lb \ebsql (s_2), db \rb) \}
&\text{subs., \repe}\\ 
 & = \{x \mapsto y ~|~ x \in \eb \lb s_1, \rep(db) \rb) \wedge 
  y \in  \eb \lb s_2, \rep(db) \rb) \}
&\text{I.H.}\\
& = \eb \lb s_1 \cprod s_2, \rep(db) \rb & \eb\ \qedhere 
 \end{align*}
\end{proof}

Note that if $s_1$ and $s_2$ do not contain duplicates, the result of the query above
can not contain duplicates.

\begin{case}
$ \repe(\sqlerc \lb \ebsql (s_1 \setminus s_2), db \rb) = 
  \eb \lb s_1 \setminus s_2, \rep(db) \rb $
\end{case}

\begin{proof}
 \begin{align*}
 &  \repe(\sqlerc \lb \ebsql (s_1 \setminus s_2), db \rb) \\
  &  = \repe(\sqlerc \lb \mbox{\sql{select} s1tmp.refkey} \\
 & ~~~~~ \sql{from } \ebsql (s_1) \mbox{ s1tmp} \\
 & ~~~~~ \mbox{\sql{where} s1tmp.refkey \sql{not in} (}\\
 & ~~~~~~~~\mbox{\sql{select} s2tmp.refkey} \\
 & ~~~~~~~~\sql{from } \ebsql(s_2) \mbox{ s2tmp}), db  \rb) 
   &\text{\ebsql}\\
 \end{align*}
 \begin{align*}
 & = \repe(\{\mbox{res}_1 ~|~ \mbox{res}_1 :  d_1 \#  \wedge \\
 & ~~~~~ \exists \mbox{s1tmp} : \sqlerc \lb \ebsql (s_1), db \rb \cdot \\
 & ~~~~~~~~  \mbox{res}_1 = \sqlerc \lb \mbox{s1tmp.refkey}, db \rb \wedge \\
 & ~~~~~~~~ \sqlerc \lb \mbox{s1tmp.refkey \sql{not in} (} \\
 & ~~~~~~~~~~~ \mbox{\sql{select} s2tmp.refkey}\\
 & ~~~~~~~~~~~ \sql{from }
   \ebsql(s_2) \mbox{ s2tmp}), db  \rb \})&\text{\sqlerc}\\
 & = \repe(\{\mbox{res}_1 ~|~ \mbox{res}_1 :  d_1 \#  \wedge \\
 & ~~~~~ \exists \mbox{s1tmp} : \sqlerc \lb \ebsql (s_1), db \rb \cdot \\
 & ~~~~~~~~  \mbox{res}_1 = \mbox{refkey(s1tmp)} \wedge \\
 & ~~~~~~~~ \neg \exists \mbox{s2tmp} : \sqlerc \lb \ebsql(s_2), db \rb \cdot \\
 & ~~~~~~~~~~~ \sqlerc \lb \mbox{s1tmp.refkey = s2tmp.refkey}, db  \rb \})
  &\text{\sqlerc}\\
 & = \repe(\{\mbox{res}_1 ~|~ \mbox{res}_1 :  d_1 \#  \wedge \\
 & ~~~~~ \exists \mbox{s1tmp} : \sqlerc \lb \ebsql (s_1), db \rb \cdot \\
 & ~~~~~~~~  \mbox{res}_1 = \mbox{refkey(s1tmp)} \wedge \\
 & ~~~~~~~~ \neg \exists \mbox{s2tmp} : \sqlerc \lb \ebsql(s_2), db \rb \cdot \\
 & ~~~~~~~~~~~ \mbox{refkey(s1tmp) = refkey(s2tmp)} \})
  &\text{\sqlerc}\\
   & = \{x ~|~ x \in  \repe(\sqlerc \lb \ebsql (s_1), db \rb) \wedge  \\
 & ~~~~~~~~~  \neg  x \in \repe(\sqlerc \lb \ebsql(s_2), db \rb)  \}
  &\text{subs., \repe}\\
     & = \{x ~|~ x \in  \eb \lb \ebsql (s_1), \rep(db) \rb \wedge  \\
 & ~~~~~~~~~~   x \not \in \eb  \lb \ebsql(s_2), \rep(db) \rb  \}
  &\text{I.H., \eb}\\
       & = \eb \lb s_1 \setminus s_2, \rep(db) \rb
  &\text{\eb~\qedhere}
 \end{align*}
\end{proof}

Note that if $s_1$ does not contain duplicates, the result of the query above
can not contain duplicates.  MySQL does not have the ANSI SQL standard
\sql{except} operator, so \ebsql\ translates $\setminus$ as shown above.

\begin{case}
$ \repe(\sqlerc \lb \ebsql (s_1 = s_2), db \rb) \equiv \eb \lb s_1 = s_2, \rep(db) \rb$
\end{case}
\begin{proof}
The proof is by Cases~\ref{case-and},~\ref{case-eq},~\ref{case-card}, 
 and~\ref{case-inter}.
\end{proof}

\begin{case}
$ \repe(\sqlerc \lb \ebsql (s_1 \subseteq s_2), db \rb) \equiv 
  \eb \lb s_1 \subseteq s_2, \rep(db) \rb$
\label{case-subseteq}
\end{case}
\begin{proof}
The proof is by Cases~\ref{case-eq},~\ref{case-card}, and~\ref{case-inter}.
\end{proof}

\begin{case}
$ \repe(\sqlerc \lb \ebsql (s_1 \subset s_2), db \rb) \equiv 
  \eb \lb s_1 \subset s_2, \rep(db) \rb $
\end{case}

\begin{proof}
The proof is by 
Cases~\ref{case-and},~\ref{case-not},~\ref{case-eq},~\ref{case-card},
and~\ref{case-inter}.
\end{proof}

\begin{case}
$ \repe(\sqlerc \lb \ebsql (x \in s), db \rb) \equiv \eb \lb x \in s, \rep(db) \rb$
\end{case}

\begin{proof}
The proof follows Case~\ref{case-subseteq}.
\end{proof}

\newpage

\begin{case}
 $ \repe (\sqlerc \lb \ebsql (\dom(r)), db \rb) = \eb \lb \dom(r), \rep(db) \rb  $
 \label{case-dom}
 \end{case}
 
\begin{proof}
 \begin{align*}
 &  \repe(\sqlerc \lb \ebsql (\dom(r)), db \rb) \\
 &  = \repe (\sqlerc \lb \sql{select distinct } \mbox{rtmp.id} & \\
 & ~~~~~~~~~~~~~~~~~~~~~~\sql{from }  \ebsql (r) \mbox{ rtmp}, db \rb) 
   &\text{\ebsql}\\
 & = \repe (\{ \mbox{res}_1 ~|~ \mbox{res}_1 :  d_1 \# \wedge \\
 & ~~~~~ \exists \mbox{rtmp} : \sqlerc \lb \ebsql (r), db \rb  \cdot & \\
 & ~~~~~~ \mbox{res}_1  = \sqlerc \lb \mbox{rtmp.id}, db \rb \})
  &\text{\sqlerc}\\
   & = \repe (\{ \mbox{res}_1 ~|~ \mbox{res}_1 :  d_1 \# \wedge \\
 & ~~~~~ \exists \mbox{rtmp} : \sqlerc \lb \ebsql (r), db \rb  \cdot & \\
 & ~~~~~~ \mbox{res}_1  = \mbox{id(rtmp)} \})
  &\text{\sqlerc}\\
   & = \{ x ~|~ x \mapsto y \in \repe(\sqlerc \lb \ebsql (r), db \rb)  \}
  &\text{subs., \repe}\\  
   & = \{ x ~|~ x \mapsto y \in \eb \lb r, \rep(db) \rb  \}
  &\text{I.H.}\\
  & = \eb \lb \dom(r), \rep(db) \rb  & \eb\ \qedhere
 \end{align*}
\end{proof}

\begin{case}
 $\repe(\sqlerc \lb \ebsql (\ran(r)), db \rb) = \eb \lb \ran(r), \rep(db) \rb$
 \end{case}
 
 \begin{proof}
The proof is analogous with Case~\ref{case-dom}.
\end{proof}

\begin{case}
$\repe(\sqlerc \lb \ebsql (s \domres r), db \rb) = \eb \lb s \domres r, \rep(db) \rb $
\label{case-domres}
\end{case}

\begin{proof}
 \begin{align*}
 &  \repe(\sqlerc \lb \ebsql (s \domres r), db \rb \\
 &  = \repe(\sqlerc \lb \mbox{\sql{select} rtmp.id, rtmp.value} \\
 & ~~~~~ \sql{from } \ebsql (r) \mbox{ rtmp},
  \ebsql(s) \mbox{ stmp} & \\
 & ~~~~~ \mbox{\sql{where} rtmp.id = stmp.refkey}, db  \rb) 
   &\text{\ebsql}\\
 & = \repe(\{\mbox{res}_1, \mbox{res}_2 ~|~ \mbox{res}_1 :  d_1 \# \wedge 
    \mbox{res}_2 :  d_2 \# \wedge\\
 & ~~~~~ \exists \mbox{rtmp} : \sqlerc \lb \ebsql (r), db \rb \cdot \\
 & ~~~~~~~~ \exists \mbox{stmp} : \sqlerc \lb \ebsql (s), db \rb \cdot \\
 & ~~~~~~~~~~~ \mbox{res}_1 = \sqlerc \lb \mbox{rtmp.id}, db \rb \wedge 
 \mbox{res}_2 = \sqlerc \lb \mbox{rtmp.value}, db \rb \wedge & \\
 & ~~~~~~~~~~~ \sqlerc \lb \mbox{rtmp.id = stmp.refkey}, db \rb \})  &\text{\sqlerc}\\
  & = \repe(\{\mbox{res}_1, \mbox{res}_2 ~|~ \mbox{res}_1 :  d_1 \# \wedge 
    \mbox{res}_2 :  d_2 \# \wedge\\
 & ~~~~~ \exists \mbox{rtmp} : \sqlerc \lb \ebsql (r), db \rb \cdot \\
 & ~~~~~~~~ \exists \mbox{stmp} : \sqlerc \lb \ebsql (s), db \rb \cdot \\
 & ~~~~~~~~~~~ \mbox{res}_1 =  \mbox{id(rtmp)} \wedge 
    \mbox{res}_2 = \mbox{value(rtmp)} \wedge & \\
 & ~~~~~~~~~~~ \mbox{id(rtmp) = refkey(stmp)} \})  &\text{\sqlerc}\\
   & = \{x \mapsto y ~|~ x \mapsto y \in \repe(\sqlerc \lb \ebsql (r), db \rb)  \\
 & ~~~~~~~~ \wedge x \in  \repe(\sqlerc \lb \ebsql (s), db \rb)  \})
    &\text{subs., \repe}\\
 & = \{x \mapsto y ~|~ x \mapsto y \in \eb \lb r, \rep(db) \rb 
 \wedge x \in  \eb \lb s, \rep(db) \rb  \})
    &\text{I.H.}\\
 &= \eb \lb s \domres r, \rep(db) \rb & \eb\ \qedhere
 \end{align*}
\end{proof}

Note that if $s$ does not contain duplicates, then the id of each tuple in $r$ matches at
most one element of $s$.  Hence, if $r$ does not contain duplicates, the result of
the query above does not contain duplicates.

\begin{case}
$\repe (\sqlerc \lb \ebsql (s \domsub r), db \rb = \eb \lb s \domsub r, \rep(db) \rb$
\label{case-domsub}
\end{case}

\begin{proof}
 \begin{align*}
 &  \repe(\sqlerc \lb \ebsql (s \domsub r), db \rb \\
  &  = \repe(\sqlerc \lb \mbox{\sql{select} rtmp.id, rtmp.value} \\
 & ~~~~~ \sql{from } \ebsql (r) \mbox{ rtmp} \\
 & ~~~~~ \mbox{\sql{where} rtmp.id \sql{not in}} \\
 & ~~~~~~~~ \mbox{(\sql{select} stmp.refkey } & \\
 & ~~~~~~~~~\sql{from } \ebsql(s) \mbox{ stmp)}, db  \rb) 
   &\text{\ebsql}\\
 & = \repe(\{\mbox{res}_1, \mbox{res}_2 ~|~ \mbox{res}_1 :  d_1 \# \wedge 
    \mbox{res}_2 :  d_2 \# \wedge\\
 & ~~~~~ \exists \mbox{rtmp} : \sqlerc \lb \ebsql (r), db \rb \cdot \\
 & ~~~~~~~~ \mbox{res}_1 = \sqlerc \lb \mbox{rtmp.id}, db \rb \wedge \\
 & ~~~~~~~~  \mbox{res}_2 = \sqlerc \lb \mbox{rtmp.value}, db \rb \wedge \\
 & ~~~~~~~~  \sqlerc \lb  \mbox{rtmp.id \sql{not in}} \\
 & ~~~~~~~~~~~ \mbox{(\sql{select} stmp.refkey} & \\
 & ~~~~~~~~~~~~ \sql{from } \ebsql(s) \mbox{stmp)}, db  \rb \}
   &\text{\sqlerc}\\
 & = \repe(\{\mbox{res}_1, \mbox{res}_2 ~|~ \mbox{res}_1 :  d_1 \# \wedge 
    \mbox{res}_2 :  d_2 \# \wedge\\
 & ~~~~~ \exists \mbox{rtmp} : \sqlerc \lb \ebsql (r), db \rb \cdot \\
 & ~~~~~~~~ \mbox{res}_1 = \mbox{id(rtmp)} \wedge
  \mbox{res}_2 = \mbox{value(rtmp)} \wedge \\
 & ~~~~~~~~  \neg \exists \mbox{stmp} :  \sqlerc \lb \ebsql(s), db \rb \cdot & \\
 & ~~~~~~~~~~~  \sqlerc \lb \mbox{rtmp.id = stmp.refkey}, db  \rb \})
   &\text{\sqlerc}\\
 & = \{x \mapsto y ~|~ x \mapsto y \in \repe(\sqlerc \lb \ebsql (r), db \rb) \wedge \\
 & ~~~~~~~~ x \not \in \repe(\sqlerc \lb \ebsql(s), db \rb) \}
   &\text{subs., \repe}\\
  & = \{x \mapsto y ~|~ x \mapsto y \in \eb \lb r, \rep(db) \rb \wedge \\
 & ~~~~~~~~ x \not \in \eb \lb s, \rep(db) \rb \}
   &\text{I.H.}\\
    & = \eb \lb s \domsub r, \rep(db) \rb  & \eb\ \qedhere
 \end{align*}
\end{proof}

Note that if $r$ does not contain duplicates, the result of the query above can
not contain duplicates.

\begin{case}
$\repe(\sqlerc \lb \ebsql (r \ranres s), db \rb) = \eb \lb r \ranres s, \rep(db) \rb$
\end{case}

\begin{proof}
The proof is analogous with Case~\ref{case-domres}.
\end{proof}

\begin{case}
$\repe(\sqlerc \lb \ebsql (r \ransub s), db \rb) = \eb \lb r \ransub s, \rep(db) \rb $
\end{case}

\begin{proof}
The proof is analogous with Case~\ref{case-domsub}.
\end{proof}

\begin{case}
$\repe(\sqlerc \lb \ebsql (r_1 \ovl r_2), db \rb) = \eb \lb r_1 \ovl r_2, \rep(db) \rb$
\end{case}

\begin{proof}
The proof is by Cases~\ref{case-union}, \ref{case-dom} and \ref{case-domsub}.
\end{proof}

\begin{case}
 $\repe(\sqlerc \lb \ebsql (r_1;r_2), db \rb) = \eb \lb r_1;r_2, \rep(db) \rb$
 \label{case-fcomp}
\end{case}

\begin{proof}
 \begin{align*}
 &  \repe(\sqlerc \lb \ebsql (r_1;r_2), db \rb) \\
  &  = \repe(\sqlerc \lb \mbox{\sql{select distinct} r1tmp.id, r2tmp.value} \\
 & ~~~~~ \sql{from } \ebsql (r_1) \mbox{ r1tmp, } \ebsql (r_2) \mbox{ r2tmp} &  \\
 & ~~~~~ \mbox{\sql{where} r1tmp.value = r2tmp.id }, db  \rb) 
   &\text{\ebsql}\\
   & = \repe(\{\mbox{res}_1, \mbox{res}_2 ~|~ \mbox{res}_1 :  d_1 \# \wedge 
    \mbox{res}_2 :  d_2 \# \wedge\\
 & ~~~~~ \exists \mbox{r1tmp} : \sqlerc \lb \ebsql (r_1), db \rb \cdot \\ 
 & ~~~~~~~~ \exists \mbox{r2tmp} : \sqlerc \lb \ebsql (r_2), db \rb \cdot \\
 & ~~~~~~~~~~~ \mbox{res}_1 = \sqlerc \lb \mbox{r1tmp.id}, db \rb \wedge \\
 & ~~~~~~~~~~~ \mbox{res}_2 = \sqlerc \lb \mbox{r2tmp.value}, db \rb \wedge 
 & \\
 & ~~~~~~~~~~~ \sqlerc \lb \mbox{r1tmp.value = r2tmp.id}, db \rb \})
  &\text{\sqlerc}\\
     & = \repe(\{\mbox{res}_1, \mbox{res}_2 ~|~ \mbox{res}_1 :  d_1 \# \wedge 
    \mbox{res}_2 :  d_2 \# \wedge\\
 & ~~~~~ \exists \mbox{r1tmp} : \sqlerc \lb \ebsql (r_1), db \rb \cdot \\ 
 & ~~~~~~~~ \exists \mbox{r2tmp} : \sqlerc \lb \ebsql (r_2), db \rb \cdot \\
 & ~~~~~~~~~~~ \mbox{res}_1 = \mbox{id(r1tmp)} \wedge 
  \mbox{res}_2 = \mbox{value(r2tmp)} \wedge  & \\
 & ~~~~~~~~~~~ \mbox{value(r1tmp) = id(r2tmp)} \})
  &\text{\sqlerc}\\
  & = \{x \mapsto y ~|~ \exists z \cdot x \mapsto z \in
 \repe(\sqlerc \lb \ebsql (r_1), db \rb) \wedge \\ 
 & ~~~~~~~~ z \mapsto y \in \repe(\sqlerc \lb \ebsql (r_2), db \rb)  \}
  &\text{subs., \repe}\\
  & = \{x \mapsto y ~|~ \exists z \cdot x \mapsto z \in
 \eb \lb r_1, \rep(db) \rb \wedge \\ 
 & ~~~~~~~~ z \mapsto y \in \eb \lb r_2, \rep(db) \rb  \}
  &\text{I.H.}\\ 
    & = \eb \lb r_1;r_2, \rep(db) \rb  & \eb\ \qedhere
 \end{align*}
\end{proof}

\begin{case}
 $\repe(\sqlerc \lb \ebsql (r_1 \circ r_2), db \rb) = \eb \lb r_1 \circ r_2, \rep(db) \rb$
\end{case}

\begin{proof}
The proof is by Case~\ref{case-fcomp}.
\end{proof}

\newpage

\begin{case}
$\repe(\sqlerc \lb \ebsql (r^{-1}), db \rb) = \eb \lb r^{-1}, \rep(db) \rb$
\end{case}

\begin{proof}
 \begin{align*}
 &  \repe(\sqlerc \lb \ebsql (r^{-1}), db \rb) \\
 & = \repe(\sqlerc \lb \sql{select } \mbox{rtmp.value, rtmp.id} & \\
 & ~~~~~ \sql{from }  \ebsql (r) \mbox{ rtmp}, db \rb)
  &\text{\ebsql}\\
 & = \repe(\{\mbox{res}_1, \mbox{res}_2 ~|~ \mbox{res}_1 :  d_1 \# \wedge 
    \mbox{res}_2 :  d_2 \# \wedge\\
 & ~~~~~ \exists \mbox{rtmp} : \sqlerc \lb \ebsql (r), db \rb \cdot  \\
 & ~~~~~~~~ \mbox{res}_1 = \sqlerc \lb \mbox{rtmp.value}, db \rb \wedge 
 & \\
 & ~~~~~~~~ \mbox{res}_2 = \sqlerc \lb \mbox{rtmp.id}, db \rb \})
  &\text{\sqlerc}\\
  & = \repe(\{\mbox{res}_1, \mbox{res}_2 ~|~ \mbox{res}_1 :  d_1 \# \wedge 
    \mbox{res}_2 :  d_2 \# \wedge\\
 & ~~~~~ \exists \mbox{rtmp} : \sqlerc \lb \ebsql (r), db \rb \cdot  \\
 & ~~~~~~~~ \mbox{res}_1 =  \mbox{value(rtmp)} \wedge 
 \mbox{res}_2 = \mbox{id(rtmp)} \})
  &\text{\sqlerc}\\
  & = \{y \mapsto x ~|~ x \mapsto y \in \repe(\sqlerc \lb \ebsql (r), db \rb) \}
  &\text{subs., \repe}\\
  & = \{y \mapsto x ~|~ x \mapsto y \in \eb \lb r, \rep(db) \rb \}
  &\text{I.H.}\\
    & = \eb \lb r^{-1}, \rep(db) \rb & \eb\ \qedhere
 \end{align*}
\end{proof}

\begin{case}
$\repe(\sqlerc \lb \ebsql (r[s]), db \rb) = \eb \lb r[s], \rep(db) \rb$
\label{case-apply}
\end{case}

\begin{proof}
 \begin{align*}
 &  \repe(\sqlerc \lb \ebsql (r[s]), db \rb) \\
 & = \repe(\sqlerc \lb \mbox{\sql{select distinct} rtmp.value}\\
 & ~~~~~\sql{from } \ebsql (r) \mbox{ rtmp, } \ebsql(s) \mbox{ stmp} & \\
 & ~~~~~\mbox{\sql{where} rtmp.id = stmp.refkey}, db \rb)
  &\text{\ebsql}\\
 & = \repe(\{\mbox{res}_1 ~|~ \mbox{res}_1 :  d_1 \#  \wedge\\
 & ~~~~~ \exists \mbox{rtmp} : \sqlerc \lb \ebsql (r), db \rb \cdot \\ 
 & ~~~~~~~~ \exists \mbox{stmp} : \sqlerc \lb \ebsql (s), db \rb \cdot \\
 & ~~~~~~~~~~~ \mbox{res}_1 = \sqlerc \lb \mbox{rtmp.value}, db \rb \wedge & 
 \\
 & ~~~~~~~~~~~  \sqlerc \lb \mbox{rtmp.id = stmp.refkey}, db \rb \})
  &\text{\sqlerc}\\
 & = \repe(\{\mbox{res}_1 ~|~ \mbox{res}_1 :  d_1 \#  \wedge\\
 & ~~~~~ \exists \mbox{rtmp} : \sqlerc \lb \ebsql (r), db \rb \cdot \\ 
 & ~~~~~~~~ \exists \mbox{stmp} : \sqlerc \lb \ebsql (s), db \rb \cdot \\
 & ~~~~~~~~~~~ \mbox{res}_1 = \mbox{value(rtmp)} \wedge 
   \mbox{id(rtmp) = refkey(stmp)} \})
  &\text{\sqlerc}\\
  & = \{y ~|~ \exists x \cdot x \in \repe(\sqlerc \lb \ebsql (s), db \rb) \wedge \\
& ~~~~~ x \mapsto y \in \repe(\sqlerc \lb \ebsql (r), db \rb)  \}
  &\text{subs., \repe}\\
    & = \{y ~|~ \exists x \cdot x \in \eb \lb s, \rep(db) \rb \wedge \\
& ~~~~~ x \mapsto y \in \eb \lb r, \rep(db) \rb  \}
  &\text{I.H.}\\
    & = \eb \lb r[s], \rep(db) \rb & \eb\ \qedhere
 \end{align*}
\end{proof}

%
%
   
\subsection{Proof of Soundness: Translating Multiple Assignments}
\label{sect-multassign}

\begin{thm}
 For every set of Event-B assignment statements $AS$ such that each
 assignment $A \in AS$ is translated directly
 to one or more SQL \textup{\sql{insert}} and \textup{\sql{delete}}
 statements 
 by the EventB2SQL tool and every correctly typed database state $db$:
\[
  \rep(\ebsqlr \lb AS, db \rb) = \ebas \lb AS, \rep(db) \rb \] 
\label{theorem-ass}
 \end{thm}
 That is, executing the database update statements generated by EventB2SQL
 from a set of Event-B assignment statements
 produces a database state that is equivalent to the Event-B model state
 produced by evaluating those assignment statements.
 The \ebas\ operator was defined in Figure~\ref{fig-eb}, and \ebsqlr\
 in Figure~\ref{fig-eb2sqla}.

  The proof proceeds by induction over the set of assignment statements.
 In the basis, $\rep(\ebsqlr \lb \lambda, db \rb) =
 \rep(\ebsqlas \lb \lambda, 
 \ebsqlos \lb \lambda, db \rb \rb) = 
 \rep(db) = \ebas \lb \lambda, \rep(db) \rb$. The inductive step is
 again done by case analysis,
 with one case for each rule for \ebsqla\ in Figure~\ref{fig-eb2sqla}.

\begin{case}
\[ \rep( \ebsqlr \lb \textup{s} \bcmeq \textup{s} \bunion s_1 || AS, db \rb)
 = 
 \ebas \lb \textup{s} \bcmeq \textup{s} \bunion s_1 || AS, \rep(db) \rb \]
\label{case-uniona}
\end{case}

\begin{proof}
\begin{align*}
& \rep( \ebsqlr \lb \mbox{s} \bcmeq \mbox{s} \bunion s_1 || AS, db \rb) \\
& =  \rep(\ebsqlas \lb \mbox{s} \bcmeq \mbox{s} \bunion s_1 || AS,
\ebsqlos \lb \mbox{s} \bcmeq \mbox{s} \bunion s_1 || AS, db \rb \rb)
 & \ebsqlr \\
& =  \rep(\ebsqlas \lb AS, [s' \rightarrow \mathit{undef}]\sqlerca \lb \\
& ~~~~~
 \ebsqla(\mbox{s} \bcmeq \mbox{s} \bunion s_1), & \ebsqlas \\ 
& ~~~~~
\ebsqlos \lb AS, \ebsqlo \lb \mbox{s} \bcmeq \mbox{s}
  \bunion s_1,  db \rb \rb \rb \rb) 
 & \ebsqlos \\ 
& =  \rep(\ebsqlas \lb AS, [s' \rightarrow \mathit{undef}]\sqlerca \lb \\
& ~~~~~
 \sql{insert ignore into} \mbox{ s } 
\sql{select } \mbox{stmp.refkey}
 \sql{ from } s' \mbox{ stmp}, & \ebsqla \\ 
& ~~~~~
\ebsqlos \lb AS, [s' \rightarrow \sqlerc \lb \ebsql(s_1),  db \rb ] db 
  \rb \rb \rb ) 
 & \ebsqlo \\ 
& =  \rep(\ebsqlas \lb AS, [s' \rightarrow \mathit{undef}]
 [\mbox{s} \rightarrow db'(\mbox{s}) \bunion \sqlerc \lb\\ 
 & ~~~~~
 \sql{select } \mbox{stmp.refkey} 
 \sql{ from } s' \mbox{ stmp}, db' \rb ] db' \rb) &
  \sqlerca, \dagger \\ 
& =  \rep(\ebsqlas \lb AS, [s' \rightarrow \mathit{undef}]
 [\mbox{s} \rightarrow db'(\mbox{s}) \bunion  \\
& ~~~~~
 \{\mbox{res}_1 | \mbox{res}_1 : d_1 \# \wedge
 \exists \mbox{stmp} : \sqlerc \lb s', db' \rb \cdot\\
& ~~~~~~~~
   \mbox{res}_1 = \sqlerc \lb \mbox{stmp.refkey}, db' \rb\}]db' \rb) &
  \sqlerc \\ 
& =  \rep(\ebsqlas \lb AS, [s' \rightarrow \mathit{undef}]
 [\mbox{s} \rightarrow db'(\mbox{s}) \bunion  \\
& ~~~~~
 \{\mbox{refkey(stmp)} |
 \mbox{stmp} \in \sqlerc \lb \ebsql(s_1), db \rb 
   \}]db' \rb) &
  \text{subs.} \\ 
& =  \rep(\ebsqlas \lb AS, 
  \ebsqlos \lb AS, 
  [\mbox{s} \rightarrow db(\mbox{s}) \bunion  \\
& ~~~~~
 \{\mbox{refkey(stmp)} |
 \mbox{stmp} \in \sqlerc \lb \ebsql(s_1), db \rb 
   \}]db \rb \rb) & * \\ 
\end{align*}
\begin{align*}
& =  \ebas \lb AS, \rep(
  [\mbox{s} \rightarrow db(\mbox{s}) \bunion  
 \sqlerc \lb \ebsql(E), db \rb 
   ]db) \rb & \text{I.H., subs.} \\ 
& =  \ebas \lb AS, 
  [\mbox{s} \rightarrow \repe(db(\mbox{s}) \bunion
  \sqlerc \lb \ebsql(E), db \rb )]\rep(db) \rb & \rep \\ 
& =  \ebas \lb AS, 
  [\mbox{s} \rightarrow \repe(db(\mbox{s})) \bunion
  \repe(\sqlerc \lb \ebsql(E), db \rb )]\rep(db) \rb & \repe \\ 
& =  \ebas \lb AS, 
  [\mbox{s} \rightarrow \rep(db)(\mbox{s}) \bunion
  \repe(\sqlerc \lb \ebsql(E), db \rb )]\rep(db) \rb & \repe \\ 
& =  \ebas \lb AS, 
  [\mbox{s} \rightarrow \eb \lb \mbox{s}, \rep(db) \rb
 \bunion
 \eb \lb E, \rep(db) \rb]\rep(db) \rb & \eb, \text{Theorem~\ref{theorem}} \\ 
& =  \ebas \lb AS, 
  [\mbox{s} \rightarrow \eb \lb \mbox{s}
 \bunion E, \rep(db) \rb]\rep(db) \rb & \eb \\ 
& =  \ebas \lb \mbox{s} \bcmeq \mbox{s} \bunion E
 || AS, \rep(db) \rb & \ebas\ \qedhere \\ 
\end{align*}
\end{proof}

In step $\dagger$, define $db' = \ebsqlos \lb AS, 
 [s' \rightarrow \sqlerc \lb \ebsql(s_1),  db \rb ] db \rb $.
In step *, note that:
\begin{itemize}
\item because an Event-B event can assign a machine variable at most once,
  no statement in $AS$ assigns to s
\item no assignment statement in $AS$ refers to $s'$
\item $db'(\mbox{s}) = db(\mbox{s})$
\item the critical parts of the right hand side of each assignment statement 
 are evaluated by \ebsqlo\ in the initial database state
 and the results stored in the state returned by
 \ebsqlos, bound to a primed version of the identifier on the left hand
 side of the assignment.  These values are then used by the SQL \sql{insert}
 and \sql{delete} statements generated for each assignment by \ebsqla.
 The net effect is to evaluate the right hand side of each assignment in
 the initial database state, preserving the simultaneous assignment
 semantics of Event-B.
\end{itemize}

\begin{case}
\[
 \rep(\ebsqlr \lb \textup{s} \bcmeq \textup{s} \setminus s_1 || AS, db \rb)
 = 
\\
 \ebas \lb \textup{s} \bcmeq \textup{s} \setminus s_1 || AS, \rep(db) \rb
\]
\end{case}

\begin{proof}
\begin{align*}
& \rep(\ebsqlr \lb \mbox{s} \bcmeq \mbox{s} \setminus s_1 || AS, db \rb) \\
& =  \rep(\ebsqlas \lb \mbox{s} \bcmeq \mbox{s} \setminus s_1 || AS, 
\ebsqlos \lb \mbox{s} \bcmeq \mbox{s} \setminus s_1 || AS, db \rb \rb) 
& \ebsqlr\\
& =  \rep(\ebsqlas \lb AS, [s' \rightarrow \mathit{undef}]\sqlerca \lb \\
& ~~~~~
 \ebsqla(\mbox{s} \bcmeq \mbox{s} \setminus s_1), & \ebsqlas \\ 
& ~~~~~
\ebsqlos \lb AS, \ebsqlo \lb \mbox{s} \bcmeq \mbox{s}
  \setminus s_1,  db \rb \rb \rb \rb) 
 & \ebsqlos \\ 
& =  \rep(\ebsqlas \lb AS, [s' \rightarrow \mathit{undef}]\sqlerca \lb \\
& ~~~~~
 \sql{delete from } \mbox{s} \sql{ where } \mbox{s.refkey} \sql{ in } \\
& ~~~~~~~~
 (\sql{select } \mbox{s1tmp.refkey } \sql{from } s' \mbox{ s1tmp})
, & \ebsqla \\ 
& ~~~~~
\ebsqlos \lb AS, [s' \rightarrow \sqlerc \lb \ebsql(s_1),  db \rb]db \rb \rb \rb) 
 & \ebsqlo \\ 
\end{align*}
\begin{align*}
& =  \rep(\ebsqlas \lb AS, [s' \rightarrow \mathit{undef}]
[\mbox{s} \rightarrow db'(\mbox{s}) \setminus
\sqlerc \lb \\
& ~~~~~
 \sql{select } \mbox{stmp.refkey} \sql{ from } \mbox{s stmp} \sql{ where }
 \mbox{stmp.refkey} \sql{ in } \\
& ~~~~~~~~
 (\sql{select } \mbox{s1tmp.refkey } \sql{from } s' \mbox{ s1tmp}),
 db' \rb]db' \rb)    
 & \sqlerca, \dagger \\ 
& =  \rep(\ebsqlas \lb AS, [s' \rightarrow \mathit{undef}]
[\mbox{s} \rightarrow db'(\mbox{s}) \setminus \\
& ~~~~~
\{\mbox{res}_1 ~|~ \mbox{res}_1 : d_1 \# \wedge
 \exists \mbox{stmp} : \sqlerc \lb \mbox{s}, db' \rb \cdot \\
& ~~~~~~~~
 \sqlerc \lb \mbox{res}_1 = \mbox{stmp.refkey}, db' \rb \wedge
 \sqlerc \lb \mbox{stmp.refkey} \sql{ in } \\
& ~~~~~~~~~~~
 (\sql{select } \mbox{s1tmp.refkey } \sql{from } s' \mbox{ s1tmp}),
 db' \rb\}]db' \rb)    
 & \sqlerc \\ 
& =  \rep(\ebsqlas \lb AS, [s' \rightarrow \mathit{undef}]
[\mbox{s} \rightarrow db'(\mbox{s}) \setminus \\
& ~~~~~
\{\mbox{refkey(stmp)} ~|~ \mbox{stmp} : db'(\mbox{s}) \wedge 
 \exists \mbox{s1tmp} : \sqlerc \lb s', db' \rb \cdot & \text{subs.,}\\
& ~~~~~~~~
 \sqlerc \lb \mbox{stmp.refkey} = \mbox{s1tmp.refkey },
 db' \rb\}]db' \rb)    
 & \sqlerc \\ 
& =  \rep(\ebsqlas \lb AS, [s' \rightarrow \mathit{undef}]
[\mbox{s} \rightarrow db'(\mbox{s}) \setminus \\
& ~~~~~
\{\mbox{refkey(stmp)} ~|~ \mbox{stmp} : db'(\mbox{s}) \wedge 
 \exists \mbox{s1tmp} : db'(s') \cdot \\
& ~~~~~~~~
 \mbox{refkey(stmp)} = \mbox{refkey(s1tmp)}
 \}]db' \rb)    
 & \sqlerc \\ 
& =  \rep(\ebsqlas \lb AS, [s' \rightarrow \mathit{undef}]
[\mbox{s} \rightarrow db'(\mbox{s}) \setminus \\
& ~~~~~
\{\mbox{refkey(stmp)} ~|~ \mbox{stmp} : db'(\mbox{s}) \wedge  \\
& ~~~~~~~~
  \mbox{stmp} : \sqlerc \lb \ebsql(s_1), db' \rb
 \}]db' \rb)    
 & \text{subs.} \\ 
& =  \rep(\ebsqlas \lb AS, 
 \ebsqlos \lb AS, 
[\mbox{s} \rightarrow db(\mbox{s}) \setminus \\
& ~~~~~
\{\mbox{refkey(stmp)} ~|~ \mbox{stmp} : db(\mbox{s}) \wedge 
  \mbox{stmp} : \sqlerc \lb \ebsql(s_1), db \rb
 \}]db \rb \rb)    
 & *\\ 
& =  \ebas \lb AS, \rep([\mbox{s} \rightarrow db(\mbox{s}) \setminus 
(db(\mbox{s}) \binter 
  \sqlerc \lb \ebsql(s_1), db \rb
 )]db) \rb
 & \text{I.H., subs.}\\ 
& =  \ebas \lb AS, [\mbox{s} \rightarrow \repe(db(\mbox{s}) \setminus 
(db(\mbox{s}) \binter \sqlerc \lb \ebsql(s_1), db \rb))]\rep(db) \rb
 & \rep\\ 
& =  \ebas \lb AS, [\mbox{s} \rightarrow \repe(db(\mbox{s})) \setminus \\
& ~~~~~
(\repe(db(\mbox{s})) \binter \repe(\sqlerc \lb \ebsql(s_1), db \rb))]\rep(db) \rb
 & \repe\\ 
& =  \ebas \lb AS, [\mbox{s} \rightarrow \rep(db)(\mbox{s}) \setminus
 & \rep \\
& ~~~~~
(\rep(db)(\mbox{s}) \binter \eb \lb s_1, \rep(db) \rb)]\rep(db) \rb
 & \text{Theorem~\ref{theorem}}\\ 
& =  \ebas \lb AS, [\mbox{s} \rightarrow \eb \lb \mbox{s}, \rep(db) \rb 
 \setminus\\
& ~~~~~
(\eb \lb \mbox{s}, \rep(db) \rb \binter \eb \lb s_1, \rep(db) \rb)]\rep(db) \rb
 & \eb\\ 
& =  \ebas \lb AS, [\mbox{s} \rightarrow \eb \lb \mbox{s}
 \setminus (\mbox{s} \binter s_1), \rep(db) \rb]\rep(db) \rb
 & \eb\\ 
& =  \ebas \lb AS, [\mbox{s} \rightarrow \eb \lb \mbox{s}
 \setminus s_1, \rep(db) \rb]\rep(db) \rb
 & **\\ 
& =  \ebas \lb \mbox{s} \bcmeq \mbox{s} \setminus s_1 || AS, 
  \rep(db) \rb
 & \ebas~\qedhere\\ 
\end{align*}
\end{proof}

In step $\dagger$, define $db' =
 \ebsqlos \lb AS, [s' \rightarrow \sqlerc \lb \ebsql(s_1),  db \rb]db
 \rb$.
In step *, see the justification for step * in Case~\ref{case-uniona}.
In step **, note that 
$s_1 \setminus (s_1 \binter s_2) = s_1 \setminus s_2$. 

\begin{case}
\[
 \rep(\ebsqlr \lb \textup{s} \bcmeq \textup{s} \binter s_1 || AS, db \rb)
 = 
 \ebas \lb \textup{s} \bcmeq \textup{s} \setminus s_1 || AS, \rep(db) \rb
\]
\end{case}

\begin{proof}
\begin{align*}
& \rep(\ebsqlr \lb \mbox{s} \bcmeq \mbox{s} \binter s_1 || AS, db \rb)\\
& =  \rep(\ebsqlas \lb \mbox{s} \bcmeq \mbox{s} \binter s_1 || AS, 
\ebsqlos \lb \mbox{s} \bcmeq \mbox{s} \binter s_1 || AS, db \rb \rb)
 & \ebsqlr \\
& =  \rep(\ebsqlas \lb AS, [\mbox{s'} \rightarrow \mathit{undef}]
  \sqlerca \lb \\
& ~~~~~
\ebsqla(\mbox{s} \bcmeq \mbox{s} \binter s_1), & \ebsqlas \\
& ~~~~~
\ebsqlos \lb AS, \ebsqlo \lb \mbox{s} 
 \bcmeq \mbox{s} \binter s_1, db \rb \rb \rb \rb) & \ebsqlos \\
& =  \rep(\ebsqlas \lb AS, [\mbox{s'} \rightarrow \mathit{undef}]
  \sqlerca \lb \\
& ~~~~~
\sql{delete from } \mbox{s} \sql{ where }
 \mbox{s.refkey } \sql{in}\\
& ~~~~~~~~ \sql{(select } \mbox{s1tmp.refkey } \sql{from }
 \mbox{s' s1tmp)}, & \ebsqla\\
& ~~~~~
\ebsqlos \lb AS, 
[\mbox{s'} \rightarrow \sqlerc \lb \ebsql(\mbox{s} \setminus s_1), 
  db \rb] db \rb \rb \rb) & \ebsqlo \\
& =  \rep(\ebsqlas \lb AS, [\mbox{s'} \rightarrow \mathit{undef}]
 [\mbox{s} \rightarrow db'(\mbox{s}) \setminus \sqlerc \lb \\
& ~~~~~
\sql{select } \mbox{stmp.refkey} \sql{ from } \mbox{s stmp} \sql{ where }
 \mbox{stmp.refkey } \sql{in}\\
& ~~~~~~~~ \sql{(select } \mbox{s1tmp.refkey } \sql{from }
 \mbox{s' s1tmp)}, db' \rb] db' \rb ) & \sqlerca, \dagger\\
& =  \rep(\ebsqlas \lb AS, [\mbox{s'} \rightarrow \mathit{undef}]
 [\mbox{s} \rightarrow db'(\mbox{s}) \setminus \\
& ~~~~~
 \{ \mbox{res}_1 ~|~ \mbox{res}_1 : d_1 \# \wedge \exists \mbox{stmp} :
 \sqlerc \lb \mbox{s}, db' \rb \cdot \\
& ~~~~~~~~
 \sqlerc \lb \mbox{res}_1 = \mbox{stmp.refkey}, db' \rb \wedge \sqlerc \lb
 \mbox{stmp.refkey } \sql{in}\\
& ~~~~~~~~~~~ \sql{(select } \mbox{s1tmp.refkey } \sql{from }
 \mbox{s' s1tmp)}, db' \rb\}] db' \rb ) & \sqlerc\\
& =  \rep(\ebsqlas \lb AS, [\mbox{s'} \rightarrow \mathit{undef}]
 [\mbox{s} \rightarrow db'(\mbox{s}) \setminus \\
& ~~~~~~~~
 \{ \mbox{refkey(stmp)} ~|~  \mbox{stmp} : db'(\mbox{s}) \wedge
 \exists \mbox{s1tmp} : \sqlerc \lb \mbox{s'}, db' \rb \cdot & \text{subs.,}\\
& ~~~~~~~~~~~
 \sqlerc \lb \mbox{stmp.refkey = s1tmp.refkey}, db' \rb\}] db' \rb ) & 
 \sqlerc\\
& =  \rep(\ebsqlas \lb AS, [\mbox{s'} \rightarrow \mathit{undef}]
 [\mbox{s} \rightarrow db'(\mbox{s}) \setminus & \text{subs.,}\\
& ~~~~~
 \{ \mbox{refkey(stmp)} ~|~  \mbox{stmp} : db'(\mbox{s}) \wedge
 \mbox{stmp} : \sqlerc \lb \mbox{s} \setminus s_1, db' \rb 
 \}] db' \rb ) & 
 \sqlerc\\
& =  \rep(\ebsqlas \lb AS, 
 [\mbox{s} \rightarrow db(\mbox{s}) \setminus \\
& ~~~~~
 \{ \mbox{refkey(stmp)} ~|~  \mbox{stmp} : db(\mbox{s}) \wedge
 \mbox{stmp} : \sqlerc \lb \mbox{s} \setminus s_1, db \rb 
 \}] db \rb ) & *\\
& =  \ebas \lb AS,  
 \rep([\mbox{s} \rightarrow db(\mbox{s}) \setminus 
 (db(\mbox{s}) \binter
  \sqlerc \lb \mbox{s} \setminus s_1, db \rb 
 )] db) \rb  & \text{I.H., subs.}\\
& =  \ebas \lb AS,  
 [\mbox{s} \rightarrow \repe(db(\mbox{s}) \setminus 
 (db(\mbox{s}) \binter
  \sqlerc \lb \mbox{s} \setminus s_1, db \rb 
 ))] \rep(db) \rb  & \rep\\
& =  \ebas \lb AS,  
 [\mbox{s} \rightarrow \repe(db(\mbox{s})) \setminus \\
& ~~~~~
 \repe(db(\mbox{s})) \binter
  \repe(\sqlerc \lb \mbox{s} \setminus s_1, db \rb 
 )] \rep(db) \rb  & \repe\\
& =  \ebas \lb AS,  
 [\mbox{s} \rightarrow \rep(db)(\mbox{s}) \setminus & \rep,\\
& ~~~~~
 \rep(db)(\mbox{s}) \binter
  \eb \lb \mbox{s} \setminus s_1, \rep(db) \rb 
 ] \rep(db) \rb  & \text{Theorem~\ref{theorem}}\\
& =  \ebas \lb AS,  
 [\mbox{s} \rightarrow \eb \lb \mbox{s} \setminus 
 (\mbox{s} \binter
  (\mbox{s} \setminus s_1)), \rep(db) \rb 
 ] \rep(db) \rb  & \eb\\
& =  \ebas \lb AS,  
 [\mbox{s} \rightarrow \eb \lb \mbox{s} \setminus 
  (\mbox{s} \setminus s_1), \rep(db) \rb 
 ] \rep(db) \rb  & **\\
& =  \ebas \lb AS,  
 [\mbox{s} \rightarrow \eb \lb \mbox{s} 
  \binter s_1, \rep(db) \rb 
 ] \rep(db) \rb  & **\\
& =  \ebas \lb \mbox{s} \bcmeq \mbox{s} \binter s_1 ||  AS,  
  \rep(db) \rb  & \ebas\ \qedhere\\
\end{align*}
\end{proof}

In step $\dagger$, define $db' = 
\ebsqlos \lb AS, 
[\mbox{s'} \rightarrow \sqlerc \lb \ebsql(\mbox{s} \setminus s_1), 
  db \rb] db \rb$.
In step *, see the justification for step * in Case~\ref{case-uniona}.
The steps marked ** are by simple set identities.

\begin{case}
\[
 \rep(\ebsqlr \lb \textup{r} \bcmeq \textup{r} \ovl r_1 || AS, db \rb)
 = 
 \ebas \lb \textup{r} \bcmeq \textup{r} \ovl r_1 || AS, \rep(db) \rb
\]
\label{case-ovla}
\end{case}

\begin{proof}
\begin{align*}
& \rep(\ebsqlr \lb \mbox{r} \bcmeq \mbox{r} \ovl r_1 || AS, db \rb) \\
& = \rep(\ebsqlas \lb \mbox{r} \bcmeq \mbox{r} \ovl r_1 || AS, 
\ebsqlos \lb \mbox{r} \bcmeq \mbox{r} \ovl r_1 || AS, db \rb \rb) 
 & \ebsqlr \\
& =  \rep(\ebsqlas \lb AS, [\mbox{r'} \rightarrow \mathit{undef}]
  \sqlerca \lb \\
& ~~~~~
 \ebsqla(\mbox{r} \bcmeq \mbox{r} \ovl r_1), & \ebsqlas\\ 
& ~~~~~
\ebsqlos \lb AS, \ebsqlo \lb \mbox{r} \bcmeq \mbox{r} \ovl r_1,  
 db \rb \rb \rb \rb) & \ebsqlos \\
& =  \rep(\ebsqlas \lb AS, [\mbox{r'} \rightarrow \mathit{undef}]
  \sqlerca \lb \\
& ~~~~~
\sql{delete from } \mbox{r} \sql{ where } \mbox{r.id} \sql{ in } \\
& ~~~~~~~~
\sql{(select } \mbox{r1tmp.id} \sql{ from } \mbox{r' r1tmp} \sql{);} \\
& ~~~~~
\sql{insert ignore into } \mbox{r}\\
& ~~~~~~~~
 \sql{ select } \mbox{r2tmp.id, r2tmp.value}
\sql{ from } \mbox{r' r2tmp} \sql{;}
, & \ebsqla\\ 
& ~~~~~
\ebsqlos \lb AS, [\mbox{r'} \rightarrow \sqlerc \lb 
 \ebsql(r_1), db \rb] 
 db \rb \rb \rb) & \sqlerco \\
& =  \rep(\ebsqlas \lb AS, [\mbox{r'} \rightarrow \mathit{undef}]
  \sqlerca \lb \\
& ~~~~~
\sql{insert ignore into } \mbox{r}\\
& ~~~~~~~~
 \sql{ select } \mbox{r2tmp.id, r2tmp.value}
\sql{ from } \mbox{r' r2tmp}, \\ 
& ~~~~~
  \sqlerca \lb 
\sql{delete from } \mbox{r} \sql{ where } \mbox{r.id} \sql{ in } \\
& ~~~~~~~~
\sql{(select } \mbox{r1tmp.id} \sql{ from } \mbox{r' r1tmp} \sql{)}, \\
& ~~~~~
\ebsqlos \lb AS, [\mbox{r'} \rightarrow \sqlerc \lb 
 \ebsql(r_1), db \rb] 
 db \rb \rb \rb \rb) & \sqlerca \\
& =  \rep(\ebsqlas \lb AS, [\mbox{r'} \rightarrow \mathit{undef}]
  \sqlerca \lb \\
& ~~~~~
\sql{insert ignore into } \mbox{r}\\
& ~~~~~~~~
 \sql{ select } \mbox{r2tmp.id, r2tmp.value}
\sql{ from } \mbox{r' r2tmp} \sql{}
, \\ 
& ~~~~~
 [\mbox{r} \rightarrow db(\mbox{r}) \setminus 
  \sqlerc \lb 
\sql{select } \mbox{rtmp.id, rtmp.value} \sql{ from } \mbox{r rtmp} \\
& ~~~~~~~~
 \sql{where } \mbox{rtmp.id} \sql{ in } \\
& ~~~~~~~~~~~
\sql{(select } \mbox{r1tmp.id} \sql{ from } \mbox{r' r1tmp} \sql{)}, 
 db' \rb] db' 
 \rb \rb) & \sqlerca, \dagger \\
\end{align*}
\begin{align*}
& =  \rep(\ebsqlas \lb AS, [\mbox{r'} \rightarrow \mathit{undef}]
  \sqlerca \lb \\
& ~~~~~
\sql{insert ignore into } \mbox{r}\\
& ~~~~~~~~
 \sql{ select } \mbox{r2tmp.id, r2tmp.value}
\sql{ from } \mbox{r' r2tmp} \sql{}
, \\ 
& ~~~~~
 [\mbox{r} \rightarrow db(\mbox{r}) \setminus 
\{ \mbox{res}_1, \mbox{res}_2 ~|~ \mbox{res}_1 : d_1 \# \wedge
 \mbox{res}_2 : d_2 \wedge \\
& ~~~~~~~~
\exists \mbox{rtmp} : 
  \sqlerc \lb \mbox{r}, db' \rb \cdot
 \sqlerc \lb \mbox{res}_1 = \mbox{rtmp.id}, db' \rb \wedge \\
& ~~~~~~~~~~~
 \sqlerc \lb \mbox{res}_2 = \mbox{rtmp.value}, db' \rb \wedge
 \sqlerc \lb  
 \mbox{rtmp.id} \sql{ in } \\
& ~~~~~~~~~~~
\sql{(select } \mbox{r1tmp.id} \sql{ from } \mbox{r' r1tmp} \sql{)}, 
 db' \rb\}] db' 
 \rb \rb) & \sqlerc \\
& =  \rep(\ebsqlas \lb AS, [\mbox{r'} \rightarrow \mathit{undef}]
  \sqlerca \lb \\
& ~~~~~
\sql{insert ignore into } \mbox{r}\\
& ~~~~~~~~
 \sql{ select } \mbox{r2tmp.id, r2tmp.value}
\sql{ from } \mbox{r' r2tmp} 
, \\ 
& ~~~~~
 [\mbox{r} \rightarrow db(\mbox{r}) \setminus 
\{ \mbox{id(rtmp), value(rtmp)} ~|~ 
 \mbox{rtmp} : db'(\mbox{r}) \wedge & \text{subs.,}\\
& ~~~~~~~~
 \exists \mbox{r1tmp} : \sqlerc \lb \mbox{r'}, db' \rb \cdot
 \sqlerc \lb \mbox{rtmp.id = r1tmp.id}, db' \rb 
 \}] db' 
 \rb \rb) & \sqlerc \\
& =  \rep(\ebsqlas \lb AS, [\mbox{r'} \rightarrow \mathit{undef}]
  \sqlerca \lb \\
& ~~~~~
\sql{insert ignore into } \mbox{r}\\
& ~~~~~~~~
 \sql{ select } \mbox{r2tmp.id, r2tmp.value}
\sql{ from } \mbox{r' r2tmp} 
, \\ 
& ~~~~~
 [\mbox{r} \rightarrow db(\mbox{r}) \setminus 
\{ \mbox{id(rtmp), value(rtmp)} ~|~ 
 \mbox{rtmp} : db'(\mbox{r}) \wedge \\
& ~~~~~~~~
 \exists \mbox{r1tmp} : db'(\mbox{r'}) \cdot
 \mbox{id(rtmp) = id(r1tmp)} 
 \}] db' 
 \rb \rb) & \sqlerc \\
& =  \rep(\ebsqlas \lb AS, [\mbox{r'} \rightarrow \mathit{undef}]
  \sqlerca \lb \\
& ~~~~~
\sql{insert ignore into } \mbox{r}\\
& ~~~~~~~~
 \sql{ select } \mbox{r2tmp.id, r2tmp.value}
\sql{ from } \mbox{r' r2tmp} 
, \\ 
& ~~~~~
 [\mbox{r} \rightarrow db(\mbox{r}) \setminus 
\{ \mbox{rtmp} ~|~ 
 \mbox{rtmp} : db(\mbox{r}) \wedge \\
& ~~~~~~~~
  \mbox{id(rtmp)} \in \sqlerc \lb \ebsql(r_1), db \rb
 \}] db' \rb \rb) & \text{subs.} \\
& =  \rep(\ebsqlas \lb AS, [\mbox{r'} \rightarrow \mathit{undef}] \\
& ~~~~~
 [\mbox{r} \rightarrow db''(\mbox{r}) \bunion \sqlerc \lb\\
& ~~~~~~~~
 \sql{ select } \mbox{r2tmp.id, r2tmp.value}
\sql{ from } \mbox{r' r2tmp}, db'' \rb] db''\rb ) &
 \ddagger, \sqlerca \\
& =  \rep(\ebsqlas \lb AS, [\mbox{r'} \rightarrow \mathit{undef}] \\
& ~~~~~
 [\mbox{r} \rightarrow db''(\mbox{r}) \bunion \{\mbox{res}_1, \mbox{res}_2 ~|~ 
\mbox{res}_1 : d_1 \# \wedge \mbox{res}_2 : d_2 \# \wedge \\
& ~~~~~~~~
 \exists \mbox{r2tmp} :
  \sqlerc \lb \mbox{r'}, db'' \rb \cdot
 \sqlerc \lb \mbox{res}_1 = \mbox{r2tmp.id}, db'' \rb \wedge \\
& ~~~~~~~~~~~
 \sqlerc \lb \mbox{res}_2 = \mbox{r2tmp.value}, db'' \rb 
 \}] db''\rb ) & \sqlerc \\
& =  \rep(\ebsqlas \lb AS, [\mbox{r'} \rightarrow \mathit{undef}]
 & \text{subs.,} \\
& ~~~~~
 [\mbox{r} \rightarrow db''(\mbox{r}) \bunion \{\mbox{id(r2tmp)}, 
 \mbox{value(r2tmp)} ~|~ 
  \mbox{r2tmp} : db''(\mbox{r'}) 
 \}] db''\rb ) & \sqlerc \\
& =  \rep(\ebsqlas \lb AS, [\mbox{r'} \rightarrow \mathit{undef}]
 [\mbox{r} \rightarrow db''(\mbox{r}) \bunion \{\mbox{id(r2tmp)}, 
 \mbox{value(r2tmp)} ~|~ \\
& ~~~~~
  \mbox{r2tmp} : \sqlerc \lb \ebsql(r_1), db \rb
 \}] db''\rb ) & \text{subs.} \\
& =  \rep(\ebsqlas \lb AS, [\mbox{r'} \rightarrow \mathit{undef}] 
 [\mbox{r} \rightarrow (db(\mbox{r}) \setminus \\
& ~~~~~
\{ \mbox{rtmp} ~|~ \mbox{rtmp} : db(\mbox{r}) \wedge 
  \mbox{id(rtmp)} \in \sqlerc \lb \ebsql(r_1), db \rb
 \}) \\
& ~~~~~~~~
 \bunion \sqlerc \lb \ebsql(r_1), db \rb
 ] db''\rb ) & \text{subs.} \\
\end{align*}
\begin{align*}
& =  \rep(\ebsqlas \lb AS, 
 [\mbox{r} \rightarrow (db(\mbox{r}) \setminus \\
& ~~~~~
\{ \mbox{rtmp} ~|~ \mbox{rtmp} : db(\mbox{r}) \wedge 
  \mbox{id(rtmp)} \in \sqlerc \lb \ebsql(r_1), db \rb
 \}) \\
& ~~~~~~~~
 \bunion \sqlerc \lb \ebsql(r_1), db \rb
 ] db \rb ) & * \\
& =  \ebas \lb AS, 
\rep(
 [\mbox{r} \rightarrow (db(\mbox{r}) \setminus \\
& ~~~~~
\{ \mbox{rtmp} ~|~ \mbox{rtmp} : db(\mbox{r}) \wedge 
  \mbox{id(rtmp)} \in \sqlerc \lb \ebsql(r_1), db \rb
 \}) \\
& ~~~~~~~~
 \bunion \sqlerc \lb \ebsql(r_1), db \rb
 ] db) \rb  & \text{I.H.} \\
& =  \ebas \lb AS, 
 [\mbox{r} \rightarrow (\repe(db(\mbox{r})) \setminus \\
& ~~~~~
\repe(\{ \mbox{rtmp} ~|~ \mbox{rtmp} : db(\mbox{r}) \wedge 
  \mbox{id(rtmp)} \in \sqlerc \lb \ebsql(r_1), db \rb
 \})) \\
& ~~~~~~~~
 \bunion \repe(\sqlerc \lb \ebsql(r_1), db \rb)
 ] \rep(db) \rb  & \rep \\
& =  \ebas \lb AS, 
 [\mbox{r} \rightarrow (\rep(db)(\mbox{r}) \setminus \\
& ~~~~~
\{ x \mapsto y ~|~ x \mapsto y \in  \repe(db(\mbox{r})) \wedge 
  x \in \repe(\sqlerc \lb \ebsql(\dom(r_1)), db \rb)
 \}) & \text{subs.,}\\
& ~~~~~~~~
 \bunion \repe(\sqlerc \lb \ebsql(r_1), db \rb)
 ] \rep(db) \rb  & \rep \\
& =  \ebas \lb AS, 
 [\mbox{r} \rightarrow (\eb \lb \mbox{r}, \rep(db) \rb
\setminus \\
& ~~~~~
\{ x \mapsto y ~|~ x \mapsto y \in  \eb \lb \mbox{r}, \rep(db) \rb \wedge 
  x \in \eb \lb \dom(r_1), \rep(db) \rb
 \}) & \text{Theorem~\ref{theorem},}\\
& ~~~~~~~~
 \bunion \eb \lb r_1, \rep(db) \rb
 ] \rep(db) \rb  & \rep \\
& =  \ebas \lb AS, 
 [\mbox{r} \rightarrow (\eb \lb \mbox{r}, \rep(db) \rb
\setminus \\
& ~~~~~
 \eb \lb \dom(r_1), \rep(db) \rb \domres \eb \lb \mbox{r}, \rep(db) \rb)
 \bunion \eb \lb r_1, \rep(db) \rb
 ] \rep(db) \rb  & \eb \\
& =  \ebas \lb AS, 
 [\mbox{r} \rightarrow \eb \lb (\mbox{r}
\setminus 
 (\dom(r_1) \domres \mbox{r}))
 \bunion r_1, \rep(db) \rb
 ] \rep(db) \rb  & \eb \\
& =  \ebas \lb AS, [\mbox{r} \rightarrow \eb \lb 
 (\dom(r_1) \domsub \mbox{r})
 \bunion r_1, \rep(db) \rb
 ] \rep(db) \rb  & ** \\
& =  \ebas \lb AS, [\mbox{r} \rightarrow \eb \lb 
  \mbox{r} \ovl r_1, \rep(db) \rb
 ] \rep(db) \rb  & \eb \\
& =  \ebas \lb \mbox{r} \bcmeq \mbox{r} \ovl r_1 || AS, 
  \rep(db) \rb  & \ebas\ \qedhere \\
\end{align*}
\end{proof}

In step $\dagger$, define $db' =
 \ebsqlos \lb AS, [\mbox{r'} \rightarrow \sqlerc \lb 
 \ebsql(r_1), db \rb] db \rb $.
In step $\ddagger$, define $db'' =
 [\mbox{r} \rightarrow db(\mbox{r}) \setminus 
\{ \mbox{rtmp} ~|~ 
 \mbox{rtmp} : db(\mbox{r}) \wedge 
  \mbox{id(rtmp)} \in \sqlerc \lb \ebsql(r_1), db \rb
 \}] db' $.
In step *, see the justification for step * in Case~\ref{case-uniona}.
In step **, $r \setminus (s \domres r) = \{x \mapsto y ~|~ x \mapsto y 
 \in r \wedge x \not \in s \domres r\} =
\{x \mapsto y ~|~ x \mapsto y \in r \wedge
 (x \mapsto y \not \in r \vee x \not \in s)\} = 
\{x \mapsto y ~|~ x \mapsto y \in r \wedge x \not \in s\} = 
s \domsub r$.

\begin{case}
\[
 \rep(\ebsqlr \lb \textup{r} \bcmeq s_1 \domsub \textup{r}  || AS, db \rb)
 = 
 \ebas \lb \textup{r} \bcmeq s_1 \domsub \textup{r} || AS, \rep(db) \rb
\]
\end{case}

\begin{proof}
\begin{align*}
& \rep(\ebsqlr \lb \mbox{r} \bcmeq s_1 \domsub \mbox{r}  || AS, db \rb) \\
& = \rep(\ebsqlas \lb \mbox{r} \bcmeq s_1 \domsub \mbox{r} || AS,
\ebsqlos \lb \mbox{r} \bcmeq s_1 \domsub \mbox{r} || AS, db \rb \rb)
 & \ebsqlr \\
& =  \rep(\ebsqlas \lb AS, [\mbox{r'} \rightarrow \mathit{undef}]
 \sqlerca \lb \\
& ~~~~~ \ebsqla(\mbox{r} \bcmeq s_1 \domsub \mbox{r}), & \ebsqlas \\
& ~~~~~
\ebsqlos \lb AS,
 \ebsqlo \lb \mbox{r} \bcmeq s_1 \domsub \mbox{r}, db \rb \rb \rb \rb) & 
 \ebsqlos \\
\end{align*}
\begin{align*}
& =  \rep(\ebsqlas \lb AS, [\mbox{r'} \rightarrow \mathit{undef}]
 \sqlerca \lb \\
& ~~~~~ \sql{delete from } \mbox{r} \sql{ where } \mbox{r.id} \sql{ in}\\
& ~~~~~~~~
\sql{(select } \mbox{stmp.refkey} \sql{ from } \mbox{r' stmp}\sql{)}, 
 & \ebsqla \\
& ~~~~~
\ebsqlos \lb AS, [\mbox{r'} \rightarrow \sqlerc \lb 
 \ebsql(s), db \rb] db \rb \rb \rb) & \ebsqlo \\
& =  \rep(\ebsqlas \lb AS, [\mbox{r'} \rightarrow \mathit{undef}] \\
& ~~~~~ [\mbox{r} \rightarrow db(\mbox{r}) \setminus \sqlerc \lb
 \sql{select } \mbox{rtmp.id, rtmp.value} \sql{ from } \mbox{r rtmp}\\
& ~~~~~~~~
 \sql{ where } \mbox{rtmp.id} \sql{ in}\\
& ~~~~~~~~~~~
\sql{(select } \mbox{stmp.refkey} \sql{ from } \mbox{r' stmp}\sql{)}, db' \rb] 
  db' \rb ) & \sqlerca, \dagger\\
& =  \rep(\ebsqlas \lb AS, [\mbox{r'} \rightarrow \mathit{undef}] \\
& ~~~~~ [\mbox{r} \rightarrow db(\mbox{r}) \setminus 
\{\mbox{res}_1, \mbox{res}_2 ~|~ \mbox{res}_1 : d_1 \# \wedge
 \mbox{res}_2 : d_2 \wedge \exists \mbox{rtmp} : 
\sqlerc \lb \mbox{r}, db' \rb \cdot \\
& ~~~~~~~~
 \sqlerc \lb \mbox{res}_1 = \mbox{rtmp.id}, db' \rb \wedge
 \sqlerc \lb \mbox{res}_2 = \mbox{rtmp.value}, db' \rb \wedge \\
& ~~~~~~~~
 \sqlerc \lb \mbox{rtmp.id} \sql{ in }
\sql{(select } \mbox{stmp.refkey} \sql{ from } \mbox{r' stmp}\sql{)}, db' \rb
 \}] 
  db' \rb ) & \sqlerc\\
& =  \rep(\ebsqlas \lb AS, [\mbox{r'} \rightarrow \mathit{undef}] \\
& ~~~~~ [\mbox{r} \rightarrow db(\mbox{r}) \setminus 
\{\mbox{id(rtmp)}, \mbox{value(rtmp)} ~|~ 
  \mbox{rtmp} : db'(\mbox{r}) \wedge & \text{subs.,}\\
& ~~~~~~~~
  \exists \mbox{stmp} : \sqlerc \lb \mbox{r'}, db' \rb \cdot
 \sqlerc \lb \mbox{rtmp.id} = \mbox{stmp.refkey}, db' \rb
 \}] 
  db' \rb ) & \sqlerc\\
& =  \rep(\ebsqlas \lb AS, [\mbox{r'} \rightarrow \mathit{undef}] 
[\mbox{r} \rightarrow db(\mbox{r}) \setminus 
 & \text{subs.,}\\
& ~~~~~ 
\{\mbox{id(rtmp)}, \mbox{value(rtmp)} ~|~ 
  \mbox{rtmp} : db'(\mbox{r}) \wedge
  \mbox{id(rtmp)} : db'(\mbox{r'})
 \}] 
  db' \rb ) & \sqlerc\\
& =  \rep(\ebsqlas \lb AS, [\mbox{r'} \rightarrow \mathit{undef}] 
[\mbox{r} \rightarrow db(\mbox{r}) \setminus 
\{\mbox{id(rtmp)}, \mbox{value(rtmp)} ~|~ \\
& ~~~~~ 
  \mbox{rtmp} : db(\mbox{r}) \wedge 
  \mbox{id(rtmp)} : \sqlerc \lb \ebsql(s_1), db \rb \}] db' \rb ) & \text{subs.}\\
& =  \rep(\ebsqlas \lb AS, 
[\mbox{r} \rightarrow db(\mbox{r}) \setminus 
\{\mbox{id(rtmp)}, \mbox{value(rtmp)} ~|~ \\
& ~~~~~ 
  \mbox{rtmp} : db(\mbox{r}) \wedge
  \mbox{id(rtmp)} : \sqlerc \lb \ebsql(s_1), db \rb \}] db \rb ) & *\\
& =  \ebas \lb AS, 
 \rep([\mbox{r} \rightarrow db(\mbox{r}) \setminus 
\{\mbox{id(rtmp)}, \mbox{value(rtmp)} ~|~ 
  \mbox{rtmp} : db(\mbox{r}) \wedge\\
& ~~~~~
  \mbox{id(rtmp)} : \sqlerc \lb \ebsql(s_1), db \rb \}] db) \rb  & \text{I.H.}\\
& =  \ebas \lb AS, 
 [\mbox{r} \rightarrow \rep(db)(r) \setminus \{x \mapsto y ~|~ 
  & \text{subs.,}\\
& ~~~~~ 
 x \mapsto y \in
  \rep(db)(\mbox{r}) \wedge
  x \in \repe(\sqlerc \lb \ebsql(s_1), db \rb)
 \}] \rep(db) \rb  & \rep\\
& =  \ebas \lb AS, 
 [\mbox{r} \rightarrow \eb \lb \rep(db)(\mbox{r}) \setminus \{x \mapsto y ~|~ 
  & \text{subs.,}\\
& ~~~~~ 
 x \mapsto y \in
  \rep(db)(\mbox{r}) \wedge
  x \in \repe(\sqlerc \lb \ebsql(s_1), db \rb)
 \}] \rep(db) \rb  & \rep\\
& =  \ebas \lb AS, 
 [\mbox{r} \rightarrow \eb \lb \mbox{r}, \rep(db) \rb
 \setminus \{x \mapsto y ~|~ 
  & \eb,\\
& ~~~~~ 
 x \mapsto y \in \eb \lb \mbox{r}, \rep(db) \rb
   \wedge
  x \in \eb \lb s_1, \rep(db) \rb
 \}] \rep(db) \rb  & \text{Theorem~\ref{theorem}}\\
& =  \ebas \lb AS, 
 [\mbox{r} \rightarrow \eb \lb \mbox{r}, \rep(db) \rb
 \setminus (\eb \lb s_1, \rep(db) \rb \domres
  \eb \lb \mbox{r}, \rep(db) \rb)
 ] \rep(db) \rb  & \eb\\
& =  \ebas \lb AS, 
 [\mbox{r} \rightarrow \eb \lb \mbox{r}
 \setminus (s \domres \mbox{r}), \rep(db) \rb
 ] \rep(db) \rb  & \eb\\
& =  \ebas \lb AS, 
 [\mbox{r} \rightarrow \eb \lb 
 s_1 \domsub \mbox{r}, \rep(db) \rb
 ] \rep(db) \rb  & **\\
& =  \ebas \lb \mbox{r} \bcmeq s_1 \domsub \mbox{r} || AS, 
 \rep(db) \rb & \ebas\ \qedhere
\end{align*}
\end{proof}

In step $\dagger$, define $db' = 
\ebsqlos \lb AS, [\mbox{r'} \rightarrow \sqlerc \lb 
 \ebsql(s_1), db \rb] db \rb$.
In step *, see the justification for step * in Case~\ref{case-uniona}.
In step **, see the justification for step ** in Case~\ref{case-ovla}.

\begin{case}
\[
 \rep(\ebsqlr \lb \textup{r} \bcmeq s_1 \domres \textup{r}  || AS, db \rb)
 = 
 \ebas \lb \textup{r} \bcmeq s_1 \domres \textup{r} || AS, \rep(db) \rb
\]
\end{case}

\begin{proof}
\begin{align*}
& \rep(\ebsqlr \lb \mbox{r} \bcmeq s_1 \domres \mbox{r}  || AS, db \rb) \\
&   \rep(\ebsqlas \lb \mbox{r} \bcmeq s_1 \domres \mbox{r} || AS,
\ebsqlos \lb \mbox{r} \bcmeq s_1 \domres \mbox{r} || AS, db \rb \rb)
 & \ebsqlr \\
& =  \rep(\ebsqlas \lb AS, [\mbox{r'} \rightarrow \mathit{undef}]
 \sqlerca \lb \\
& ~~~~~ \ebsqla(\mbox{r} \bcmeq s_1 \domres \mbox{r}), & \ebsqlas \\
& ~~~~~ \ebsqlos \lb AS, 
 \ebsqlo \lb \mbox{r} \bcmeq s_1 \domres \mbox{r}), db \rb \rb \rb \rb) &
  \ebsqlos \\
& =  \rep(\ebsqlas \lb AS, [\mbox{r'} \rightarrow \mathit{undef}]
 \sqlerca \lb \\
& ~~~~~ 
\sql{delete from } \mbox{r} \sql{ where } \mbox{r.id} \sql{ in}\\
& ~~~~~~~~
  \sql{(select } \mbox{stmp.refkey} \sql{ from } \mbox{r' stmp}\sql{)},
  & \ebsqla \\
& ~~~~~
\ebsqlos \lb AS, 
[\mbox{r'} \rightarrow \sqlerc \lb 
 \ebsql(\dom(\mbox{r}) \setminus s_1), db \rb] db  \rb \rb \rb) & \ebsqlo \\
& =  \rep(\ebsqlas \lb AS, [\mbox{r'} \rightarrow \mathit{undef}]
[\mbox{r} \rightarrow db(\mbox{r}) \setminus \\
& ~~~~~ \sqlerc \lb
\sql{select } \mbox{rtmp.id, rtmp.value} \sql{ from } \mbox{r rtmp} \\
& ~~~~~~~~
 \sql{ where } \mbox{rtmp.id} \sql{ in}\\
& ~~~~~~~~~~~
  \sql{(select } \mbox{stmp.refkey} \sql{ from } \mbox{r' stmp}\sql{)}, 
 db' \rb] db' \rb) & \sqlerca, \dagger\\
& =  \rep(\ebsqlas \lb AS, [\mbox{r'} \rightarrow \mathit{undef}]
[\mbox{r} \rightarrow db(\mbox{r}) \setminus \\
& ~~~~~ 
\{\mbox{res}_1, \mbox{res}_2 ~|~ \mbox{res}_1 : d_1 \# \wedge
\mbox{res}_2 : d_2 \# \wedge
 \exists \mbox{rtmp} : \sqlerc \lb \mbox{r}, db' \rb \cdot\\
& ~~~~~~~~
\sqlerc \lb \mbox{res}_1 = \mbox{rtmp.id}, db' \rb \wedge
\sqlerc \lb \mbox{res}_2 = \mbox{rtmp.value}, db' \rb \wedge\\
& ~~~~~~~~~~~
 \sqlerc \lb \mbox{rtmp.id} \sql{ in}\\
& ~~~~~~~~~~~~~~
  \sql{(select } \mbox{stmp.refkey} \sql{ from } \mbox{r' stmp}\sql{)}, 
 db' \rb\}] db' \rb) & \sqlerc\\
& =  \rep(\ebsqlas \lb AS, [\mbox{r'} \rightarrow \mathit{undef}]
[\mbox{r} \rightarrow db(\mbox{r}) \setminus \\
& ~~~~~ 
\{\mbox{id(rtmp)}, \mbox{value(rtmp)} ~|~ 
  \mbox{rtmp} : db'(\mbox{r}) \wedge & \text{subs.,}\\
& ~~~~~~~~
 \exists \mbox{stmp} : \sqlerc \lb \mbox{r'}, db' \rb \cdot
 \sqlerc \lb \mbox{rtmp.id = stmp.refkey},
 db' \rb\}] db' \rb) & \sqlerc\\
& =  \rep(\ebsqlas \lb AS, [\mbox{r'} \rightarrow \mathit{undef}]
[\mbox{r} \rightarrow db(\mbox{r}) \setminus \\
& ~~~~~ 
\{\mbox{id(rtmp)}, \mbox{value(rtmp)} ~|~ 
  \mbox{rtmp} : db'(\mbox{r}) \wedge  \\
& ~~~~~~~~
 \exists \mbox{stmp} : db'(\mbox{r'}) \cdot
  \mbox{id(rtmp) = refkey(stmp)}
 \}] db' \rb) & \sqlerc\\
& =  \rep(\ebsqlas \lb AS, [\mbox{r'} \rightarrow \mathit{undef}]
[\mbox{r} \rightarrow db(\mbox{r}) \setminus \\
& ~~~~~ 
\{\mbox{id(rtmp)}, \mbox{value(rtmp)} ~|~ 
  \mbox{rtmp} : db(\mbox{r}) \wedge  \\
& ~~~~~~~~
  \mbox{id(rtmp)} : \sqlerc \lb 
 \ebsql(\dom(\mbox{r}) \setminus s_1), db \rb
 \}] db' \rb) & \text{subs.}\\
& =  \rep(\ebsqlas \lb AS,
[\mbox{r} \rightarrow db(\mbox{r}) \setminus 
\{\mbox{id(rtmp)}, \mbox{value(rtmp)} ~|~ \\
& ~~~~~ 
  \mbox{rtmp} : db(\mbox{r}) \wedge 
  \mbox{id(rtmp)} : \sqlerc \lb 
 \ebsql(\dom(\mbox{r}) \setminus s_1), db \rb
 \}] db \rb) & *\\
& =  \ebas \lb AS,
\rep([\mbox{r} \rightarrow db(\mbox{r}) \setminus 
\{\mbox{id(rtmp)}, \mbox{value(rtmp)} ~|~ \\
& ~~~~~ 
  \mbox{rtmp} : db(\mbox{r}) \wedge  
  \mbox{id(rtmp)} : \sqlerc \lb 
 \ebsql(\dom(\mbox{r}) \setminus s_1), db \rb
 \}] db) \rb & \text{I.H.}\\
\end{align*}
\begin{align*}
& =  \ebas \lb AS,
[\mbox{r} \rightarrow \rep(db)(\mbox{r}) \setminus 
\repe(\{\mbox{id(rtmp)}, \mbox{value(rtmp)} ~|~ \\
& ~~~~~ 
  \mbox{rtmp} : db(\mbox{r}) \wedge
  \mbox{id(rtmp)} : \sqlerc \lb 
 \ebsql(\dom(\mbox{r}) \setminus s_1), db \rb
 \})] \rep(db) \rb & \rep\\
& =  \ebas \lb AS,
[\mbox{r} \rightarrow \rep(db)(\mbox{r}) \setminus 
\{x \mapsto y ~|~ x \mapsto y \in \rep(db)(\mbox{r}) \wedge  & \text{subs.,}\\
& ~~~~~ 
  x \in \repe(\sqlerc \lb 
 \ebsql(\dom(\mbox{r}) \setminus s_1), db \rb)
 \}] \rep(db) \rb & \rep\\
& =  \ebas \lb AS,
[\mbox{r} \rightarrow \eb \lb \mbox{r}, \rep(db) \rb
\setminus 
\{x \mapsto y ~|~ x \mapsto y \in \eb \lb \mbox{r}, \rep(db) \rb \wedge
   & \eb,\\
& ~~~~~ 
  x \in \eb \lb \dom(\mbox{r}) \setminus s_1, \rep(db) \rb
 \}] \rep(db) \rb & \text{Theorem~\ref{theorem}}\\
& =  \ebas \lb AS,
[\mbox{r} \rightarrow \eb \lb \mbox{r}, \rep(db) \rb
\setminus 
\eb \lb (\dom(\mbox{r}) \setminus s_1) \domres \mbox{r}, \rep(db) \rb
 ] \rep(db) \rb & \eb\\
& =  \ebas \lb AS,
[\mbox{r} \rightarrow \eb \lb \mbox{r}
\setminus 
((\dom(\mbox{r}) \setminus s_1) \domres \mbox{r}), \rep(db) \rb
 ] \rep(db) \rb & \eb\\
& =  \ebas \lb AS,
[\mbox{r} \rightarrow \eb \lb \mbox{r}
\setminus 
(s_1 \domsub \mbox{r}), \rep(db) \rb
 ] \rep(db) \rb & **\\
& =  \ebas \lb AS,
[\mbox{r} \rightarrow \eb \lb s_1 \domres \mbox{r}, \rep(db) \rb
 ] \rep(db) \rb & ***\\
& =  \ebas \lb \mbox{r} \bcmeq s_1 \domres \mbox{r} || AS,
  \rep(db) \rb & \ebas\ \qedhere\\
\end{align*}
\end{proof}

In step $\dagger$, define $db' =
\ebsqlos \lb AS, 
[\mbox{r'} \rightarrow \sqlerc \lb 
 \ebsql(\dom(\mbox{r}) \setminus s_1), db \rb] db  \rb$
In step *, see the justification for step * in Case~\ref{case-uniona}.
In step **, see the justification for step ** in Case~\ref{case-ovla}.
In step ***, $r \setminus (s \domsub r) =
\{x \mapsto y ~|~ x \mapsto y \in r \wedge x \mapsto y \not \in
 (s \domsub r) \} =
\{x \mapsto y ~|~ x \mapsto y \in r \wedge (x \mapsto y \not \in
 r \vee x \in s) \} =
\{x \mapsto y ~|~ x \mapsto y \in r \wedge x \in s \} = s \domres r$

\begin{case}
\[
 \rep(\ebsqlr \lb \textup{r} \bcmeq \textup{r} \ransub s_1 || AS, db \rb)
 = 
 \ebas \lb \textup{r} \bcmeq \textup{r} \ransub s_1 || AS, \rep(db) \rb
\]
\end{case}

\begin{proof}
\begin{align*}
& \rep(\ebsqlr \lb \mbox{r} \bcmeq \mbox{r} \ransub s_1 || AS, \rb) \\
& = \rep(\ebsqlas \lb \mbox{r} \bcmeq \mbox{r} \ransub s_1 || AS, 
\ebsqlos \lb \mbox{r} \bcmeq \mbox{r} \ransub s_1 || AS, db \rb \rb)
 & \ebsqlr \\
& =  \rep(\ebsqlas \lb AS, [\mbox{r'} \rightarrow \mathit{undef}]
 \sqlerca \lb\\
& ~~~~~
  \ebsqla(\mbox{r} \bcmeq \mbox{r} \ransub s_1), & \ebsqlas\\
& ~~~~~
\ebsqlos \lb AS, 
 \ebsqlo \lb \mbox{r} \bcmeq \mbox{r} \ransub s_1, db \rb \rb \rb \rb) & \ebsqlos \\
& =  \rep(\ebsqlas \lb AS, [\mbox{r'} \rightarrow \mathit{undef}]
 \sqlerca \lb\\
& ~~~~~
 \sql{delete from } \mbox{r} \sql{ where } \mbox{r.value} \sql{ in} \\
& ~~~~~~~~
\sql{(select } \mbox{stmp.refkey} \sql{ from } \mbox{r' stmp}\sql{)},
 & \ebsqla\\
& ~~~~~
\ebsqlos \lb AS, [\mbox{r'} \rightarrow \sqlerc \lb 
 \ebsql(s_1), db \rb] db \rb \rb \rb) & \ebsqlo \\
& =  \rep(\ebsqlas \lb AS, [\mbox{r'} \rightarrow \mathit{undef}]
[\mbox{r} \rightarrow db(\mbox{r}) \setminus
 \sqlerc \lb\\
& ~~~~~~~~
 \sql{select } \mbox{rtmp.id, rtmp.value} \sql{ from } \mbox{r rtmp}
\sql{ where } \mbox{rtmp.value} \sql{ in} \\
& ~~~~~~~~~~~
\sql{(select } \mbox{stmp.refkey} \sql{ from } \mbox{r' stmp}\sql{)},
 db' \rb] db' \rb)
& \sqlerca, \dagger \\
\end{align*}
\begin{align*}
& =  \rep(\ebsqlas \lb AS, [\mbox{r'} \rightarrow \mathit{undef}]\\
& ~~~~~
[\mbox{r} \rightarrow db(\mbox{r}) \setminus
 \{\mbox{res}_1, \mbox{res}_2 ~|~ \mbox{res}_1 : d_1 \# \wedge
\mbox{res}_2 : d_2 \# \wedge 
 \exists \mbox{rtmp} : \sqlerc \lb \mbox{r}, db' \rb \cdot \\
& ~~~~~~~~
 \sqlerc \lb \mbox{res}_1 = \mbox{rtmp.id}, db' \rb \wedge
 \sqlerc \lb \mbox{res}_2 = \mbox{rtmp.value}, db' \rb \wedge\\
& ~~~~~~~~
 \sqlerc \lb \mbox{rtmp.value} \sql{ in} \\
& ~~~~~~~~~~~
\sql{(select } \mbox{stmp.refkey} \sql{ from } \mbox{r' stmp}\sql{)},
 db' \rb\}] db' \rb)
& \sqlerc \\
& =  \rep(\ebsqlas \lb AS, [\mbox{r'} \rightarrow \mathit{undef}]\\
& ~~~~~
[\mbox{r} \rightarrow db(\mbox{r}) \setminus
 \{\mbox{id(rtmp)}, \mbox{value(rtmp)} ~|~ 
 \mbox{rtmp} : db'(\mbox{r}) \wedge & \text{subs.,}\\
& ~~~~~~~~
 \exists \mbox{stmp} : \sqlerc \lb \mbox{r'}, db' \rb \cdot
 \sqlerc \lb \mbox{rtmp.value} =
 \mbox{stmp.refkey},
 db' \rb\}] db' \rb)
& \sqlerc \\
& =  \rep(\ebsqlas \lb AS, [\mbox{r'} \rightarrow \mathit{undef}]
[\mbox{r} \rightarrow db(\mbox{r}) \setminus  & \text{subs.,}\\
& ~~~~~
 \{\mbox{id(rtmp)}, \mbox{value(rtmp)} ~|~ 
 \mbox{rtmp} : db'(\mbox{r}) \wedge 
 \mbox{value(rtmp)} : db'(\mbox{r'})
 \}] db' \rb)
& \sqlerc \\
& =  \rep(\ebsqlas \lb AS, 
[\mbox{r} \rightarrow db(\mbox{r}) \setminus 
 \{\mbox{id(rtmp)}, \mbox{value(rtmp)} ~|~ \\
& ~~~~~
 \mbox{rtmp} : db(\mbox{r}) \wedge 
 \mbox{value(rtmp)} :  \sqlerc \lb \ebsql(s_1), db \rb
 \}] db \rb)
& \text{subs., *} \\
& =  \ebas \lb AS, \rep([\mbox{r} \rightarrow db(\mbox{r}) \setminus 
 \{\mbox{id(rtmp)}, \mbox{value(rtmp)} ~|~ \\
& ~~~~~
 \mbox{rtmp} : db(\mbox{r}) \wedge 
 \mbox{value(rtmp)} :  \sqlerc \lb \ebsql(s_1), db \rb
 \}] db) \rb
& \text{I.H.} \\
& =  \ebas \lb AS, [\mbox{r} \rightarrow \rep(db)(\mbox{r}) \setminus 
 \repe(\{\mbox{id(rtmp)}, \mbox{value(rtmp)} ~|~ \\
& ~~~~~
 \mbox{rtmp} : db(\mbox{r}) \wedge 
 \mbox{value(rtmp)} :  \sqlerc \lb \ebsql(s_1), db \rb
 \})] \rep(db) \rb
& \rep \\
& =  \ebas \lb AS, [\mbox{r} \rightarrow \rep(db)(\mbox{r}) \setminus 
 \{x \mapsto y ~|~ 
& \text{subs.,} \\
& ~~~~~
 x \mapsto y \in \rep(db)(\mbox{r}) \wedge 
 y \in \repe(\sqlerc \lb \ebsql(s_1), db \rb)
 \}] \rep(db) \rb & \rep \\
& =  \ebas \lb AS, [\mbox{r} \rightarrow \eb \lb \mbox{r}, \rep(db) \rb 
  \setminus \{x \mapsto y ~|~ 
 & \eb, \\
& ~~~~~
 x \mapsto y \in \eb \lb \mbox{r}, \rep(db) \rb \wedge
 y \in \eb \lb s_1, \rep(db) \rb)
 \}] \rep(db) \rb & \text{Theorem~\ref{theorem}} \\
& =  \ebas \lb AS, [\mbox{r} \rightarrow \eb \lb \mbox{r}, \rep(db) \rb 
  \setminus 
 \eb \lb \mbox{r} \ranres s_1, \rep(db) \rb] \rep(db) \rb & \eb \\
& =  \ebas \lb AS, [\mbox{r} \rightarrow \eb \lb \mbox{r}
  \setminus (\mbox{r} \ranres s), \rep(db) \rb] \rep(db) \rb & \eb \\
& =  \ebas \lb AS, [\mbox{r} \rightarrow \eb \lb \mbox{r}
  \ransub s_1, \rep(db) \rb] \rep(db) \rb & ** \\
& =  \ebas \lb \mbox{r} \bcmeq \mbox{r} \ransub s_1 || AS, \rep(db) \rb
  & \ebas\ \qedhere \\
\end{align*}
\end{proof}
 
In step $\dagger$, define $db' =
 \ebsqlos \lb AS, [\mbox{r'} \rightarrow \sqlerc \lb 
 \ebsql(s_1), db \rb] db \rb$.
In step *, see the justification for step * in Case~\ref{case-uniona}.
In step **, note that $r \setminus (r \ranres s) =
\{x \mapsto y ~|~ x \mapsto y \in r \wedge x \mapsto y \not \in
 (r \ranres s) \} =
\{x \mapsto y ~|~ x \mapsto y \in r \wedge (x \mapsto y \not \in
 r \vee y \not \in s) \} =
\{x \mapsto y ~|~ x \mapsto y \in r \wedge  y \not \in s) \} =
r \ransub s$.

\begin{case}
\[
 \rep(\ebsqlr \lb \textup{r} \bcmeq \textup{r} \ranres s_1 || AS, db \rb)
 = 
 \ebas \lb \textup{r} \bcmeq \textup{r} \ranres s_1 || AS, \rep(db) \rb
\]
\end{case}

\begin{proof}
\begin{align*}
& \rep(\ebsqlr \lb \mbox{r} \bcmeq \mbox{r} \ranres s_1 || AS, db \rb) \\
& = \rep(\ebsqlas \lb \mbox{r} \bcmeq \mbox{r} \ranres s_1 || AS, 
\ebsqlos \lb \mbox{r} \bcmeq \mbox{r} \ranres s_1 || AS, db \rb \rb) 
 & \ebsqlr \\
\end{align*}
\begin{align*}
& = \rep(\sqlercas \lb \ebsqlas(AS), [\mbox{r'} \rightarrow \mathit{undef}]
 \sqlerca \lb \\
& ~~~~~
\ebsqla(\mbox{r} \bcmeq \mbox{r} \ranres s_1), & \ebsqlas\\
& ~~~~~
\ebsqlos \lb AS,  
 \ebsqlo \lb \mbox{r} \bcmeq \mbox{r} \ranres s_1,
db \rb \rb \rb \rb) & \ebsqlos\\
& = \rep(\ebsqlas \lb AS, [\mbox{r'} \rightarrow \mathit{undef}]
 \sqlerca \lb \\
& ~~~~~
\sql{delete from } \mbox{r} \sql{ where } \mbox{r.value} \sql{ in}\\
& ~~~~~~~~
\sql{(select } \mbox{stmp.refkey} \sql{ from } \mbox{r' stmp}\sql{)}, 
 & \ebsqla\\
& ~~~~~
\ebsqlos \lb AS, 
 [\mbox{r'} \rightarrow \sqlerc \lb 
 \ebsql(\ran(\mbox{r}) \setminus s_1), db \rb] db 
\rb \rb \rb) & \sqlerco\\
& = \rep(\ebsqlas \lb AS, [\mbox{r'} \rightarrow \mathit{undef}]
[\mbox{r} \rightarrow db(\mbox{r}) \setminus \\
& ~~~~~
\{\mbox{res}_1, \mbox{res}_2 ~|~ \mbox{res}_1 : d_1 \# \wedge
\mbox{res}_2 : d_2 \# \wedge
\exists \mbox{rtmp} : \sqlerc \lb \mbox{r}, db' \rb \cdot \\
& ~~~~~~~~
\sqlerc \lb \mbox{res}_1 = \mbox{rtmp.id}, db' \rb \wedge
\sqlerc \lb \mbox{res}_2 = \mbox{rtmp.value}, db' \rb \wedge \\
& ~~~~~~~~
\sqlerc \lb \mbox{rtmp.value} \sql{ in}\\
& ~~~~~~~~~~~
\sql{(select } \mbox{stmp.refkey} \sql{ from } \mbox{r' stmp}\sql{)}, 
 db' \rb\}] db' \rb)
& \sqlerca, \dagger\\
& = \rep(\ebsqlas \lb AS, [\mbox{r'} \rightarrow \mathit{undef}]
[\mbox{r} \rightarrow db(\mbox{r}) \setminus \\
& ~~~~~
\{\mbox{id(rtmp)}, \mbox{value(rtmp)} ~|~ 
\mbox{rtmp} : db'(\mbox{r}) \wedge
\exists \mbox{stmp} : \sqlerc \lb \mbox{r'}, db' \rb \cdot & \text{subs.,}\\
 & ~~~~~~~~
 \sqlerc \lb \mbox{rtmp.value} = \mbox{stmp.refkey}, db' \rb
\}] db' \rb)
& \sqlerc\\
& = \rep(\ebsqlas \lb AS, [\mbox{r'} \rightarrow \mathit{undef}]
[\mbox{r} \rightarrow db(\mbox{r}) \setminus \\
& ~~~~~
\{\mbox{id(rtmp)}, \mbox{value(rtmp)} ~|~ 
\mbox{rtmp} : db(\mbox{r}) \wedge
\exists \mbox{stmp} : db'(\mbox{r'}) \cdot & \text{subs.,}\\
 & ~~~~~~~~
 \mbox{value(rtmp)} = \mbox{refkey(stmp)}
\}] db' \rb)
& \sqlerc\\
& = \rep(\ebsqlas \lb AS, [\mbox{r'} \rightarrow \mathit{undef}]
[\mbox{r} \rightarrow db(\mbox{r}) \setminus \\
& ~~~~~
\{\mbox{id(rtmp)}, \mbox{value(rtmp)} ~|~ 
\mbox{rtmp} : db(\mbox{r}) \wedge
\exists \mbox{stmp} : \sqlerc \lb 
 \ebsql(\ran(\mbox{r}) \setminus s_1, db \rb \cdot \\
 & ~~~~~~~~
 \mbox{value(rtmp)} = \mbox{refkey(stmp)}
\}] db' \rb)
& \text{subs.}\\
& = \rep(\ebsqlas \lb AS, 
[\mbox{r} \rightarrow db(\mbox{r}) \setminus 
\{\mbox{id(rtmp)}, \mbox{value(rtmp)} ~|~ 
\mbox{rtmp} : db(\mbox{r}) \wedge \\
& ~~~~~
\exists \mbox{stmp} : \sqlerc \lb 
 \ebsql(\ran(\mbox{r}) \setminus s_1), db \rb \cdot
 \mbox{value(rtmp)} = \mbox{refkey(stmp)}
\}] db \rb)
& *\\
& = \ebas \lb AS, 
\rep([\mbox{r} \rightarrow db(\mbox{r}) \setminus 
\{\mbox{id(rtmp)}, \mbox{value(rtmp)} ~|~ 
\mbox{rtmp} : db(\mbox{r}) \wedge\\
& ~~~~~
\exists \mbox{stmp} : \sqlerc \lb 
 \ebsql(\ran(\mbox{r}) \setminus s_1), db \rb \cdot
 \mbox{value(rtmp)} = \mbox{refkey(stmp)}
\}] db) \rb
& \text{I.H.}\\
& = \ebas \lb AS, 
[\mbox{r} \rightarrow \rep(db)(\mbox{r}) \setminus 
\repe(\{\mbox{id(rtmp)}, \mbox{value(rtmp)} ~|~
\mbox{rtmp} : db(\mbox{r}) \wedge \\
& ~~~~~
\exists \mbox{stmp} : \sqlerc \lb 
 \ebsql(\ran(\mbox{r}) \setminus s_1), db \rb \cdot 
 \mbox{value(rtmp)} = \mbox{refkey(stmp)}
\})] \rep(db) \rb
& \rep\\
& = \ebas \lb AS, 
[\mbox{r} \rightarrow \rep(db)(\mbox{r}) \setminus 
\{x \mapsto y ~|~ x \mapsto y \in \rep(db)(\mbox{r}) \wedge & \text{subs.,}\\
& ~~~~~ y \in \repe(\sqlerc \lb \ebsql(\ran(\mbox{r}) \setminus s_1), db \rb)
\}] \rep(db) \rb
& \rep\\
& = \ebas \lb AS, 
[\mbox{r} \rightarrow \eb \lb \mbox{r}, \rep(db) \rb \setminus 
\{x \mapsto y ~|~ x \mapsto y \in \eb \lb \mbox{r}, \rep(db) \rb \wedge & 
 \eb,\\
& ~~~~~ y \in \eb \lb \ran(\mbox{r}) \setminus s_1, \rep(db) \rb
\}] \rep(db) \rb
& \text{Theorem~\ref{theorem}}\\
& = \ebas \lb AS, 
[\mbox{r} \rightarrow \eb \lb \mbox{r}, \rep(db) \rb \setminus 
 \eb \lb \mbox{r} \ranres (\ran(\mbox{r}) \setminus s_1), \rep(db) \rb
] \rep(db) \rb
& \eb\\
& = \ebas \lb AS, 
[\mbox{r} \rightarrow \eb \lb \mbox{r} \setminus 
 (\mbox{r} \ranres (\ran(\mbox{r}) \setminus s_1)), \rep(db) \rb
] \rep(db) \rb
& \eb\\
& = \ebas \lb AS, 
[\mbox{r} \rightarrow \eb \lb \mbox{r} \setminus 
 (\mbox{r} \ransub s_1), \rep(db) \rb
] \rep(db) \rb
& **\\
& = \ebas \lb AS, 
[\mbox{r} \rightarrow \eb \lb \mbox{r} \ranres s_1, \rep(db) \rb
] \rep(db) \rb
& ***\\
& = \ebas \lb \mbox{r} \bcmeq \mbox{r} \ranres s_1 || AS, \rep(db) \rb
& \ebas\ \qedhere\\
\end{align*}
\end{proof}

In step $\dagger$, define $db' =
\ebsqlos \lb AS, [\mbox{r'} \rightarrow \sqlerc \lb 
 \ebsql(\ran(\mbox{r}) \setminus s_1), db \rb] db \rb $.
In step *, see the justification for step * in Case~\ref{case-uniona}.
In step **, $r \ranres (\ran(r) \setminus s) =
\{x \mapsto y ~|~ x \mapsto y \in r \wedge y \in (\ran(r) \setminus s) \} =
\{x \mapsto y ~|~ x \mapsto y \in r \wedge y \in \ran(r) \wedge 
 y \not \in s \} =
\{x \mapsto y ~|~ x \mapsto y \in r \wedge  y \not \in s \} =
r \ransub s$.
In step ***, $r \setminus (r \ransub s) =
\{x \mapsto y ~|~ x \mapsto y \in r \wedge x \mapsto y \not \in
  (r \ransub s) \} =
\{x \mapsto y ~|~ x \mapsto y \in r \wedge (x \mapsto y \not \in
  r \vee y \in s) \} =
\{x \mapsto y ~|~ x \mapsto y \in r \wedge y \in s \} =
r \ranres s $.

\begin{case}
\[
 \rep( \ebsqlr \lb \textup{s} \bcmeq s_1 || AS, db \rb)
 = 
 \ebas \lb \textup{s} \bcmeq s_1 || AS, \rep(db) \rb
\]
\end{case}

\begin{proof}
\begin{align*}
& \rep( \ebsqlr \lb \mbox{s} \bcmeq s_1 || AS, db \rb) \\
& = \rep(\ebsqlas \lb \mbox{s} \bcmeq s_1 || AS,
\ebsqlos \lb \mbox{s} \bcmeq s_1 || AS, db \rb \rb) & \ebsqlr \\
& = \rep(\ebsqlas \lb AS, [\mbox{s'} \rightarrow \mathit{undef}]
 \sqlerca \lb \\
& ~~~~~
\ebsqla(\mbox{s} \bcmeq s_1), & \ebsqlas\\
& ~~~~~
\ebsqlos \lb AS,
 \ebsqlo \lb \mbox{s} \bcmeq s_1,
db \rb \rb \rb \rb) & \ebsqlos\\
& = \rep(\ebsqlas \lb AS, [\mbox{s'} \rightarrow \mathit{undef}]
 \sqlerca \lb \\
& ~~~~~
\sql{delete from } \mbox{s}\sql{;}\\
& ~~~~~
\sql{insert ignore into } \mbox{s} \sql{ select } \mbox{s1tmp.refkey}
\sql{ from } \mbox{s' s1tmp}, & \ebsqla \\
& ~~~~~
\ebsqlos \lb AS, [\mbox{s'} \rightarrow \sqlerc \lb 
 \ebsql(s_1),
db \rb] db \rb \rb \rb) & \sqlerco\\
& = \rep(\ebsqlas \lb AS, [\mbox{s'} \rightarrow \mathit{undef}]
 \sqlerca \lb \\
& ~~~~~
\sql{insert ignore into } \mbox{s} \sql{ select } \mbox{s1tmp.refkey}
\sql{ from } \mbox{s' s1tmp}, \\
& ~~~~~
\sqlerca \lb \sql{delete from } \mbox{s}, \\
& ~~~~~~~~
\ebsqlos \lb AS, [\mbox{s'} \rightarrow \sqlerc \lb 
 \ebsql(s_1),
db \rb] db \rb \rb \rb \rb) & \sqlerca\\
& = \rep(\ebsqlas \lb AS, [\mbox{s'} \rightarrow \mathit{undef}]
 \sqlerca \lb \\
& ~~~~~
\sql{insert ignore into } \mbox{s} \sql{ select } \mbox{s1tmp.refkey}
\sql{ from } \mbox{s' s1tmp}, \\
& ~~~~~
[\mbox{s} \rightarrow \{\}] 
\ebsqlos \lb AS,
[\mbox{s'} \rightarrow \sqlerc \lb 
 \ebsql(s_1),
db \rb] db \rb \rb \rb) & \sqlerca\\
& = \rep(\ebsqlas \lb AS, [\mbox{s'} \rightarrow \mathit{undef}]
 [\mbox{s} \rightarrow db'(s) \bunion \\
& ~~~~~
\sqlerc \lb \sql{select } \mbox{s1tmp.refkey}
\sql{ from } \mbox{s' s1tmp}, db' \rb ] db' \rb) & \sqlerca, \dagger\\
& = \rep(\ebsqlas \lb AS, [\mbox{s'} \rightarrow \mathit{undef}]
 [\mbox{s} \rightarrow db'(s) \bunion \\
& ~~~~~ \{\mbox{res}_1 ~|~ \mbox{res}_1 : d_1 \# \wedge
\exists \mbox{s1tmp} : \sqlerc \lb \mbox{s'}, db' \rb \cdot \\
& ~~~~~~~~ \sqlerc \lb \mbox{res}_1 =  \mbox{s1tmp.refkey}, db' \rb \}
] db' \rb) & \sqlerc\\
& = \rep(\ebsqlas \lb AS, [\mbox{s'} \rightarrow \mathit{undef}]
 [\mbox{s} \rightarrow \{\} \bunion & \text{subs.,}\\
& ~~~~~ \{\mbox{refkey(s1tmp)} ~|~ 
 \mbox{s1tmp} : db'(\mbox{s'}) \} ] db' \rb) & \sqlerc\\
& = \rep(\ebsqlas \lb AS, [\mbox{s'} \rightarrow \mathit{undef}]
 [\mbox{s} \rightarrow\\
& ~~~~~
 \{\mbox{refkey(s1tmp)} ~|~ 
 \mbox{s1tmp} : \sqlerc \lb \ebsql(s_1), db \rb \} ] db' \rb) & \text{subs.}\\
& = \rep(\ebsqlas \lb AS, [\mbox{s'} \rightarrow \mathit{undef}]
 [\mbox{s} \rightarrow
 \sqlerc \lb \ebsql(s_1), db \rb ] db' \rb) & \text{subs.}\\
& = \rep(\ebsqlas \lb AS, 
 [\mbox{s} \rightarrow
 \sqlerc \lb \ebsql(s_1), db \rb ] db \rb) & *\\
\end{align*}
\begin{align*}
& = \ebas \lb AS, 
 \rep([\mbox{s} \rightarrow
 \sqlerc \lb \ebsql(s_1), db \rb ] db) \rb & \text{I.H.}\\
& = \ebas \lb AS, 
 [\mbox{s} \rightarrow
 \repe(\sqlerc \lb \ebsql(s_1), db \rb) ] \rep(db) \rb & \rep\\
& = \ebas \lb AS, 
 [\mbox{s} \rightarrow
 \eb \lb s_1, \rep(db) \rb ] \rep(db) \rb & \text{Theorem~\ref{theorem}}\\
& = \ebas \lb \mbox{s} \bcmeq s_1 || AS, \rep(db) \rb & \ebas\ \qedhere\\
\end{align*}
\end{proof}

In step $\dagger$, define $db' =
[\mbox{s} \rightarrow \{\}] 
\ebsqlos \lb AS, 
[\mbox{s'} \rightarrow \sqlerc \lb \ebsql(s_1), db \rb] db \rb$. 
In step *, see the justification for step * in Case~\ref{case-uniona}.

\begin{case}
\[
 \rep( \ebsqlr \lb \textup{r} \bcmeq r_1 || AS, db \rb)
 = 
 \ebas \lb \textup{r} \bcmeq r_1 || AS, \rep(db) \rb
\]
\end{case}

\begin{proof}
\begin{align*}
& \rep( \ebsqlr \lb \mbox{r} \bcmeq r_1 || AS, db \rb) \\
& = \rep(\ebsqlas \lb \mbox{r} \bcmeq r_1 || AS,
\ebsqlos \lb \mbox{r} \bcmeq r_1 || AS, db \rb \rb) & \ebsqlr \\
& = \rep(\ebsqlas \lb AS, [\mbox{r'} \rightarrow \mathit{undef}]
 \sqlerca \lb \\
& ~~~~~
\ebsqla(\mbox{r} \bcmeq r_1), & \ebsqlas\\
& ~~~~~
\ebsqlos \lb AS,
 \ebsqlo \lb \mbox{r} \bcmeq r_1,
db \rb \rb \rb \rb) & \ebsqlos\\
& = \rep(\ebsqlas \lb AS, [\mbox{r'} \rightarrow \mathit{undef}]
 \sqlerca \lb \\
& ~~~~~
\sql{delete from } \mbox{r}\sql{;}\\
& ~~~~~
\sql{insert ignore into } \mbox{r} \sql{ select } \mbox{r1tmp.id, r1tmp.value}
\sql{ from } \mbox{r' r1tmp}, & \ebsqla \\
& ~~~~~
\ebsqlos \lb AS, [\mbox{r'} \rightarrow \sqlerc \lb 
 \ebsql(r_1),
db \rb] db \rb \rb \rb) & \sqlerco\\
& = \rep(\ebsqlas \lb AS, [\mbox{r'} \rightarrow \mathit{undef}]
 \sqlerca \lb \\
& ~~~~~
\sql{insert ignore into } \mbox{r} \sql{ select } \mbox{r1tmp.id, r1tmp.value}
\sql{ from } \mbox{r' r1tmp}, \\
& ~~~~~
\sqlerca \lb \sql{delete from } \mbox{r}, \\
& ~~~~~~~~
\ebsqlos \lb AS, [\mbox{r'} \rightarrow \sqlerc \lb 
 \ebsql(r_1),
db \rb] db \rb \rb \rb \rb) & \sqlerca\\
& = \rep(\ebsqlas \lb AS, [\mbox{r'} \rightarrow \mathit{undef}]
 \sqlerca \lb \\
& ~~~~~
\sql{insert ignore into } \mbox{r} \sql{ select } \mbox{r1tmp.id, rt1mp.value}
\sql{ from } \mbox{r' r1tmp}, \\
& ~~~~~
[\mbox{r} \rightarrow \{\}] 
\ebsqlos \lb AS,
[\mbox{r'} \rightarrow \sqlerc \lb 
 \ebsql(r_1),
db \rb] db \rb \rb \rb) & \sqlerca\\
& = \rep(\ebsqlas \lb AS, [\mbox{r'} \rightarrow \mathit{undef}]
 [\mbox{r} \rightarrow db'(r) \bunion \\
& ~~~~~
\sqlerc \lb \sql{select } \mbox{r1tmp.id, r1tmp.value}
\sql{ from } \mbox{r' r1tmp}, db' \rb ] db' \rb) & \sqlerca, \dagger\\
& = \rep(\ebsqlas \lb AS, [\mbox{r'} \rightarrow \mathit{undef}]
 [\mbox{r} \rightarrow db'(r) \bunion \\
& ~~~~~ \{\mbox{res}_1, \mbox{res}_2 ~|~ \mbox{res}_1 : d_1 \# \wedge
 \mbox{res}_2 : d_2 \# \wedge
\exists \mbox{r1tmp} : \sqlerc \lb \mbox{r'}, db' \rb \cdot \\
& ~~~~~~~~ \sqlerc \lb \mbox{res}_1 =  \mbox{r1tmp.id}, db' \rb
\wedge \sqlerc \lb \mbox{res}_2 =  \mbox{r1tmp.value}, db' \rb \}
] db' \rb) & \sqlerc\\
& = \rep(\ebsqlas \lb AS, [\mbox{r'} \rightarrow \mathit{undef}]
 [\mbox{r} \rightarrow \{\} \bunion & \text{subs.,}\\
& ~~~~~ \{\mbox{id(r1tmp), value(r1tmp)} ~|~ 
 \mbox{r1tmp} : db'(\mbox{r'}) \} ] db' \rb) & \sqlerc\\
& = \rep(\ebsqlas \lb AS, [\mbox{r'} \rightarrow \mathit{undef}]
 [\mbox{r} \rightarrow\\
& ~~~~~
 \{\mbox{id(r1tmp), value(r1tmp)} ~|~ 
 \mbox{r1tmp} : \sqlerc \lb \ebsql(r_1), db \rb \} ] db' \rb) & \text{subs.}\\
\end{align*}
\begin{align*}
& = \rep(\ebsqlas \lb AS, [\mbox{r'} \rightarrow \mathit{undef}]
 [\mbox{r} \rightarrow
 \sqlerc \lb \ebsql(r_1), db \rb ] db' \rb) & \text{subs.}\\
& = \rep(\ebsqlas \lb AS, 
 [\mbox{r} \rightarrow
 \sqlerc \lb \ebsql(r_1), db \rb ] db \rb) & *\\
& = \ebas \lb AS, 
 \rep([\mbox{r} \rightarrow
 \sqlerc \lb \ebsql(r_1), db \rb ] db) \rb & \text{I.H.}\\
& = \ebas \lb AS, 
 [\mbox{r} \rightarrow
 \repe(\sqlerc \lb \ebsql(r_1), db \rb) ] \rep(db) \rb & \rep\\
& = \ebas \lb AS, 
 [\mbox{r} \rightarrow
 \eb \lb r_1, \rep(db) \rb ] \rep(db) \rb & \text{Theorem~\ref{theorem}}\\
& = \ebas \lb \mbox{r} \bcmeq r_1 || AS, \rep(db) \rb & \ebas\ \qedhere\\
\end{align*}
\end{proof}

In step $\dagger$, define $db' =
[\mbox{r} \rightarrow \{\}] 
\ebsqlos \lb AS, 
[\mbox{r'} \rightarrow \sqlerc \lb \ebsql(r_1), db \rb] db \rb$. 
In step *, see the justification for step * in Case~\ref{case-uniona}.

\section{Related Work}

Soundness proofs are important for any
code generation tool.
Proof techniques vary depending on language properties and semantics, and
the availability of proof support tools.
%
For example, the correctness of a translation from a large 
subset of C to assembly language was verified in the CompCert 
project~\cite{Leroy2009}.  All verified parts of the CompCert compiler
are written in Coq~\cite{bertot2004}, as are the semantics of the source
and target 
languages.  The soundness (semantic preservation) proof was completed
with the aid of the Coq proof assistant.  This approach is closely
related to the idea of embedding a formal specification language in the
language of an automated theorem prover, which is a popular technique
for reasoning about soundness~\cite{Bowen1994,Jacquel2011}.

The authors of~\cite{Furst2014} take a correctness-by-construction approach
to verify that the code generated by their system is a refinement of the
Event-B model it was generated from.  In particular, their tool automatically
produces the invariants needed to show that the generated code is a
refinement, as well as the invariants needed for proving the well-definedness
of arithmetic operations.  These proofs can then be carried out with the
aid of automated theorem provers.  However, their tool can translate only
a limited subset of Event-B with restricted schedules, and a separate
correctness proof needs to be performed for each model that is translated.
The E-SPARK approach~\cite{murali2012} similarly translates an Event-B
model to an implementation in the SPARK subset of Ada, annotated
with the loop invariants and pre- and post-conditions needed for verification.
The verification conditions can then be discharged with SPARK proof tools.
E-SPARK translates only concrete, sequential Event-B models and all events
are merged into a single procedure in the implementation.

The CLawZ toolset~\cite{ohalloran2013} also uses a correctness-by-construction
approach.  Here, a Simulink\textsuperscript{\textregistered} model is 
translated to both a Z specification and executable code (in the SPARK subset
of Ada).  Additionally, the tool generates the verification conditions
needed to show that the code is a refinement of the Z specification.
The authors state that the Theorem Prover included in the toolset seems
to be able to discharge these verification conditions without human
interaction due to the regular structure of the generated specifications and
code.
Note that Event-B provides a much richer modeling notation than
Simulink\textsuperscript{\textregistered}.

The EB$^3$TG tool~\cite{Gervais:2009} generates database applications
(implemented in Java) from EB$^3$ specifications.  EB$^3$ is a trace-based
formal specification language -- system outputs are specified in terms 
of sequences of input events -- and so is quite different from state-based
languages such as Event-B.  The EB$^3$TG tool synthesizes a database 
transaction for each type of input event, as well as the SQL statements
for creating database tables.  The authors provide sketches of manual
proofs of both the soundness
and completeness of their code generation technique, using a pair of 
morphisms that map from EB$^3$ traces to database states, and from
database states back to traces.  Both proofs proceed by structural induction,
with the soundness proof using the morphism from traces to states to show
the appropriate correspondence, and the completeness proof the morphism
from states to traces.  The soundness proof is similar to ours in that
a mapping is used to show the correspondence of states after the evaluation
of a specification construct and the execution of code generated from that
construct.  The translation performed by EventB2SQL is not complete, as
there are Event-B constructs such as operations on infinite sets for which
EventB2SQL does not generate code.  As such, no proof of completeness is
possible.

Perhaps the most closely related work to ours is the UB2SQL 
tool~\cite{Mammar:2006}.  The authors propose a methodology for developing
information systems that begins with UML class diagrams specifying the
structure of the
data in the system, and state and collaboration diagrams modeling the
functionality.  These diagrams are automatically translated to a
number of abstract B~\cite{TheBBook} machines, which are then refined in a
 largely
automatic manner to implementation-level B machines.  Finally,
statements for creating the necessary database tables and a Java/JDBC
implementation of the operations are derived.  The authors were working
on automating the code generation step at the time the paper cited above
was published.  They state that proving the soundess of their code generation
technique is difficult because code derivation is done outside of any
formal environment, and indeed their code generation rules are presented
somewhat informally.  However, they have proven the soundness of the
refinement rules used to product the implementation-level B machines using
AtelierB~\cite{atelierb}, and have integrated tactics for each rule into the
B prover.
The key philosophical difference from our work is that UB2SQL is
intended for integrating the formality of the B method into 
the development of information systems with an explicit database 
component, while EventB2SQL is meant for generating code from any
Event-B model for which persistent state is useful.

By comparison with the approaches described above, our soundness proof is
rather straightforward.  This is due in part to the restricted scope of
the problem we are considering.  By looking only at the translation of
sets of assignment statements, we avoid issues such as 
atomicity, scheduling, \ldots that a full soundness proof would
need to consider.  Additionally, we were fortunate to find two formal
semantics for SQL~\cite{sql1994,Meira1990} that were so compatible with
the way that we had formalized our translation algorithm and the
mathematical definitions of Event-B operations.

\section{Future Work and Conclusion}

The semantics and proof presented in this work consider only Event-B
operators that \ebsql\ translates directly to SQL queries.  Several other
constructs (powerset, direct product, parallel product, partition, 
set comprehensions and quantified assertions)
are translated to a mixture of SQL and Java code.  Defining the semantics of 
this part of the translation is considerably more challenging, as is proving its
soundness.  Defining a full semantics for the translation of events presents
a similar challenge, as an event is translated to a Java method that 
contains JDBC API calls.  Finally, a machine checked version of the proof
presented in this paper would provide a substantially greater level of 
assurance of the soundness of the translation.  Given the number of languages
and translation operators involved, conducting such a proof presents a
formidable challenge.


 As noted in the introduction to this work, soundness is  critical
 for any code generator for Event-B.
 Because EventB2SQL translates abstract machines and so avoids the refinement
 chains typically seen in Event-B developments, the model and implementation
 are very different and the correspondence between them is not readily
 apparent to the user.
 Hence, it is particularly important to provide assurance that the code
 generated by the EventB2SQL tool is sound.
As the translation of sets of Event-B assignment statements presented and 
proven sound in this
work forms the core of the tool, the proof provides
that assurance, giving users confidence 
that the generated code satisfies the correctness and safety properties that
were verified for the original Event-B model.

\bibliographystyle{abbrv}
\bibliography{proof}
\end{document}